\documentclass[12pt]{article}
\input epsf

\def\Z{\mbox{$\hbox{Z}$}}
\def\W{\mbox{$\hbox{W}$}}
\def\R{\mbox{$\hbox{R}$}}

\newcommand{\gevc}{\ensuremath{{\rm GeV}\!/c}}
\newcommand{\gevcc}{\ensuremath{{\rm GeV}\!/c^2}}

\newcommand{\epem}{\ensuremath{{{\rm e}^+{\rm e}^-}}}
\newcommand{\epemto}{\ensuremath{{{\rm e}^+{\rm e}^- \to}}}
\newcommand{\invpb}{\ensuremath{{\rm pb^{-1}}}}
\newcommand{\st}{\ensuremath{{\tilde{\rm t}}}}
\newcommand{\sbt}{\ensuremath{{\tilde{\rm b}}}}
\newcommand{\sq}{\ensuremath{{\tilde{\rm q}}}}
\newcommand{\glu}{\ensuremath{{\tilde{\rm g}}}}
\newcommand{\mst}{\ensuremath{m_\st}}

\newcommand{\msq}{\ensuremath{m_\sq}}
\newcommand{\mglu}{\ensuremath{m_\glu}}
\newcommand{\mZ}{\ensuremath{m_{\rm Z}}}
\newcommand{\ststbar}{\ensuremath{\st\bar{\st}}}
\newcommand{\sbsbbar}{\ensuremath{\sbt\bar{\sbt}}}
\newcommand{\sqsqbar}{\ensuremath{\sq\bar{\sq}}}
\newcommand{\sqqbar}{\ensuremath{\sq\bar{\rm q}}}
\newcommand{\qqbar}{\ensuremath{{\rm q\bar{q}}}}
\newcommand{\bbbar}{\ensuremath{{\rm b\bar{b}}}}
\newcommand{\ccbar}{\ensuremath{{\rm c\bar{c}}}}

\newcommand{\nnbar}{\ensuremath{\nu\bar{\nu}}}
\newcommand{\ro}{\ensuremath{{\rm R}^0}}
\newcommand{\rp}{\ensuremath{{\rm R}^\pm}}

\def\Journal#1#2#3#4{{#1} {\bf #2} (#3) #4}
\def\etal{et al.}

\def\NIMA{{\em Nucl. Instrum. and Methods} {\bf A}}
\def\NPB{{\em Nucl. Phys.} {\bf  B}}

\def\PLB{{\em Phys. Lett.} {\bf  B}}

\def\PRL{\em Phys. Rev. Lett.}

\def\PRD{{\em Phys. Rev.} {\bf D}}

\def\EPJ{{\em Eur. Phys. J.} {\bf C}}
\def\PREP{\em Phys. Rept.}
\def\CPC{\em Comput. Phys. Commun.}
\def\RNC{\em Riv. Nuovo Cim.}

\font\ninerm=cmr9

\begin{document}

\pagenumbering{arabic}
\pagestyle{plain}

\date{}
\title{ \null\vspace{1cm}
Search for stable hadronizing squarks and gluinos\\
in \epem\ collisions up to $\sqrt{s} = 209$\,GeV
\vspace{1cm}}
\author{The ALEPH Collaboration$^*)$}
%\author{The ALEPH Collaboration\\
%{\footnotesize Corresponding authors: A.\,Kraan, P.\,Janot, G.\,Sguazzoni$^*)$}}

\maketitle

\begin{picture}(160,1)
\put(0,105){\rm ORGANISATION EUROP\'EENNE POUR LA RECHERCHE NUCL\'EAIRE (CERN)}
\put(30,100){\rm Laboratoire Europ\'een pour la Physique des Particules}
\put(125,84){\parbox[t]{45mm}{\tt CERN-EP/2003-024}}
\put(125,78){\parbox[t]{45mm}{\tt 12-May-2003}}
%\put(125,72){\parbox[t]{45mm}{\tt Final Draft}}
%\put(125,84){\parbox[t]{45mm}{\tt ALEPH/2003-004}}
%\put(125,78){\parbox[t]{45mm}{\tt CONF/2003-002}}
%\put(125,72){\parbox[t]{45mm}{\tt 7-Mar-2003}}
\end{picture}

\vspace{.2cm}
\begin{abstract}
\vspace{.2cm}
Searches for stable, hadronizing scalar quarks and gluinos are 
performed using the data collected with the ALEPH detector at LEP. Gluon
splitting into a gluino or a squark pair is searched for at 
centre-of-mass energies around the Z resonance, in the 
\epemto\ \qqbar\glu\glu\ and \qqbar\sqsqbar\ processes. Stable 
squark pair production, and stop pair production with 
subsequent decays into a stable gluino, $\st \to {\rm c} \glu$, are also 
directly searched for at centre-of-mass energies from 183 to 209\,GeV.

\ 

Altogether, stable hadronizing stop (sbottom) quarks are 
excluded up to masses of 95 (92)\,\gevcc, and stable hadronizing gluinos
are excluded up to 26.9\,\gevcc, at 95\% confidence level. In the framework 
of R-parity-conserving supersymmetric models in which the gluino and the 
stop quark are the two lightest supersymmetric 
particles, a 95\%\,C.L. lower limit of 80\,\gevcc\ is set on the stop 
quark mass. 

\end{abstract}

\vfill
\centerline{\it Submitted to The European Physical Journal C}
\vskip .5cm
\noindent
--------------------------------------------\hfil\break
{\ninerm $^*)$ See next pages for the list of authors}

\eject

%\vfill
%\centerline{\it Contribution to the Winter Conferences 2003}
%\vskip .5cm
%--------------------------------------------\hfil\break
%{\ninerm $^*)$ E-mail: Aafke.Kraan@cern.ch, Patrick.Janot@cern.ch,
%Giacomo.Sguazzoni@cern.ch}
%%
%\eject

%------------------------------------------------------------------------
% authob12pt.tex
% authors' list for papers at LEP 1.5 and 2 energies
%-----------------------------------------------------------------------
\pagestyle{empty}
\newpage
\small
%
% remember the old settings
%
\newlength{\saveparskip}
\newlength{\savetextheight}
\newlength{\savetopmargin}
\newlength{\savetextwidth}
\newlength{\saveoddsidemargin}
\newlength{\savetopsep}
\setlength{\saveparskip}{\parskip}
\setlength{\savetextheight}{\textheight}
\setlength{\savetopmargin}{\topmargin}
\setlength{\savetextwidth}{\textwidth}
\setlength{\saveoddsidemargin}{\oddsidemargin}
\setlength{\savetopsep}{\topsep}
%
% text dimensions for the author list
%
\setlength{\parskip}{0.0cm}
\setlength{\textheight}{25.0cm}
\setlength{\topmargin}{-1.5cm}
\setlength{\textwidth}{16 cm}
\setlength{\oddsidemargin}{-0.0cm}
\setlength{\topsep}{1mm}
\pretolerance=10000
%%\begin{document}
%\centerline{EUROPEAN ORGANIZATION FOR NUCLEAR RESEARCH}
%\centerline{EUROPEAN LABORATORY FOR PARTICLE PHYSICS (CERN)}
%\vspace{1cm}
%\begin{flushright}CERN-EP-2000-
%\7 May 2003 - last update
%\end{flushright}
\centerline{\large\bf The ALEPH Collaboration}
\footnotesize
\vspace{0.5cm}
{\raggedbottom
\begin{sloppypar}
\samepage\noindent
A.~Heister,
S.~Schael
\nopagebreak
\begin{center}
\parbox{15.5cm}{\sl\samepage
Physikalisches Institut das RWTH-Aachen, D-52056 Aachen, Germany}
\end{center}\end{sloppypar}
\vspace{2mm}
\begin{sloppypar}
\noindent
R.~Barate,
R.~Bruneli\`ere,
I.~De~Bonis,
D.~Decamp,
C.~Goy,
S.~Jezequel,
J.-P.~Lees,
F.~Martin,
E.~Merle,
\mbox{M.-N.~Minard},
B.~Pietrzyk,
B.~Trocm\'e
\nopagebreak
\begin{center}
\parbox{15.5cm}{\sl\samepage
Laboratoire de Physique des Particules (LAPP), IN$^{2}$P$^{3}$-CNRS,
F-74019 Annecy-le-Vieux Cedex, France}
\end{center}\end{sloppypar}
\vspace{2mm}
\begin{sloppypar}
\noindent
S.~Bravo,
M.P.~Casado,
M.~Chmeissani,
J.M.~Crespo,
E.~Fernandez,
M.~Fernandez-Bosman,
Ll.~Garrido,$^{15}$
M.~Martinez,
A.~Pacheco,
H.~Ruiz
\nopagebreak
\begin{center}
\parbox{15.5cm}{\sl\samepage
Institut de F\'{i}sica d'Altes Energies, Universitat Aut\`{o}noma
de Barcelona, E-08193 Bellaterra (Barcelona), Spain$^{7}$}
\end{center}\end{sloppypar}
\vspace{2mm}
\begin{sloppypar}
\noindent
A.~Colaleo,
D.~Creanza,
N.~De~Filippis,
M.~de~Palma,
G.~Iaselli,
G.~Maggi,
M.~Maggi,
S.~Nuzzo,
A.~Ranieri,
G.~Raso,$^{24}$
F.~Ruggieri,
G.~Selvaggi,
L.~Silvestris,
P.~Tempesta,
A.~Tricomi,$^{3}$
G.~Zito
\nopagebreak
\begin{center}
\parbox{15.5cm}{\sl\samepage
Dipartimento di Fisica, INFN Sezione di Bari, I-70126 Bari, Italy}
\end{center}\end{sloppypar}
\vspace{2mm}
\begin{sloppypar}
\noindent
X.~Huang,
J.~Lin,
Q. Ouyang,
T.~Wang,
Y.~Xie,
R.~Xu,
S.~Xue,
J.~Zhang,
L.~Zhang,
W.~Zhao
\nopagebreak
\begin{center}
\parbox{15.5cm}{\sl\samepage
Institute of High Energy Physics, Academia Sinica, Beijing, The People's
Republic of China$^{8}$}
\end{center}\end{sloppypar}
\vspace{2mm}
\begin{sloppypar}
\noindent
D.~Abbaneo,
T.~Barklow,$^{26}$
O.~Buchm\"uller,$^{26}$
M.~Cattaneo,
B.~Clerbaux,$^{23}$
H.~Drevermann,
R.W.~Forty,
M.~Frank,
F.~Gianotti,
J.B.~Hansen,
J.~Harvey,
D.E.~Hutchcroft,$^{30}$,
P.~Janot,
B.~Jost,
M.~Kado,$^{2}$
P.~Mato,
A.~Moutoussi,
F.~Ranjard,
L.~Rolandi,
D.~Schlatter,
G.~Sguazzoni,
F.~Teubert,
A.~Valassi,
I.~Videau
\nopagebreak
\begin{center}
\parbox{15.5cm}{\sl\samepage
European Laboratory for Particle Physics (CERN), CH-1211 Geneva 23,
Switzerland}
\end{center}\end{sloppypar}
\vspace{2mm}
\begin{sloppypar}
\noindent
F.~Badaud,
S.~Dessagne,
A.~Falvard,$^{20}$
D.~Fayolle,
P.~Gay,
J.~Jousset,
B.~Michel,
S.~Monteil,
D.~Pallin,
J.M.~Pascolo,
P.~Perret
\nopagebreak
\begin{center}
\parbox{15.5cm}{\sl\samepage
Laboratoire de Physique Corpusculaire, Universit\'e Blaise Pascal,
IN$^{2}$P$^{3}$-CNRS, Clermont-Ferrand, F-63177 Aubi\`{e}re, France}
\end{center}\end{sloppypar}
\vspace{2mm}
\begin{sloppypar}
\noindent
J.D.~Hansen,
J.R.~Hansen,
P.H.~Hansen,
A.C.~Kraan,
B.S.~Nilsson
\nopagebreak
\begin{center}
\parbox{15.5cm}{\sl\samepage
Niels Bohr Institute, 2100 Copenhagen, DK-Denmark$^{9}$}
\end{center}\end{sloppypar}
\vspace{2mm}
\begin{sloppypar}
\noindent
A.~Kyriakis,
C.~Markou,
E.~Simopoulou,
A.~Vayaki,
K.~Zachariadou
\nopagebreak
\begin{center}
\parbox{15.5cm}{\sl\samepage
Nuclear Research Center Demokritos (NRCD), GR-15310 Attiki, Greece}
\end{center}\end{sloppypar}
\vspace{2mm}
\begin{sloppypar}
\noindent
A.~Blondel,$^{12}$
\mbox{J.-C.~Brient},
F.~Machefert,
A.~Roug\'{e},
H.~Videau
\nopagebreak
\begin{center}
\parbox{15.5cm}{\sl\samepage
Laoratoire Leprince-Ringuet, Ecole
Polytechnique, IN$^{2}$P$^{3}$-CNRS, \mbox{F-91128} Palaiseau Cedex, France}
\end{center}\end{sloppypar}
\vspace{2mm}
\begin{sloppypar}
\noindent
V.~Ciulli,
E.~Focardi,
G.~Parrini
\nopagebreak
\begin{center}
\parbox{15.5cm}{\sl\samepage
Dipartimento di Fisica, Universit\`a di Firenze, INFN Sezione di Firenze,
I-50125 Firenze, Italy}
\end{center}\end{sloppypar}
\vspace{2mm}
\begin{sloppypar}
\noindent
A.~Antonelli,
M.~Antonelli,
G.~Bencivenni,
F.~Bossi,
G.~Capon,
F.~Cerutti,
V.~Chiarella,
P.~Laurelli,
G.~Mannocchi,$^{5}$
G.P.~Murtas,
L.~Passalacqua
\nopagebreak
\begin{center}
\parbox{15.5cm}{\sl\samepage
Laboratori Nazionali dell'INFN (LNF-INFN), I-00044 Frascati, Italy}
\end{center}\end{sloppypar}
\vspace{2mm}
%\pagebreak
\begin{sloppypar}
\noindent
J.~Kennedy,
J.G.~Lynch,
P.~Negus,
V.~O'Shea,
A.S.~Thompson
\nopagebreak
\begin{center}
\parbox{15.5cm}{\sl\samepage
Department of Physics and Astronomy, University of Glasgow, Glasgow G12
8QQ,United Kingdom$^{10}$}
\end{center}\end{sloppypar}
\vspace{2mm}
%\pagebreak
\begin{sloppypar}
\noindent
S.~Wasserbaech
\nopagebreak
\begin{center}
\parbox{15.5cm}{\sl\samepage
Utah Valley State College, Orem, UT 84058, U.S.A.}
\end{center}\end{sloppypar}
\vspace{2mm}
\pagebreak
\begin{sloppypar}
\noindent
R.~Cavanaugh,$^{4}$
S.~Dhamotharan,$^{21}$
C.~Geweniger,
P.~Hanke,
V.~Hepp,
E.E.~Kluge,
A.~Putzer,
H.~Stenzel,
K.~Tittel,
M.~Wunsch$^{19}$
\nopagebreak
\begin{center}
\parbox{15.5cm}{\sl\samepage
Kirchhoff-Institut f\"ur Physik, Universit\"at Heidelberg, D-69120
Heidelberg, Germany$^{16}$}
\end{center}\end{sloppypar}
\vspace{2mm}
\begin{sloppypar}
\noindent
R.~Beuselinck,
W.~Cameron,
G.~Davies,
P.J.~Dornan,
M.~Girone,$^{1}$
R.D.~Hill,
N.~Marinelli,
J.~Nowell,
S.A.~Rutherford,
J.K.~Sedgbeer,
J.C.~Thompson,$^{14}$
R.~White
\nopagebreak
\begin{center}
\parbox{15.5cm}{\sl\samepage
Department of Physics, Imperial College, London SW7 2BZ,
United Kingdom$^{10}$}
\end{center}\end{sloppypar}
\vspace{2mm}
\begin{sloppypar}
\noindent
V.M.~Ghete,
P.~Girtler,
E.~Kneringer,
D.~Kuhn,
G.~Rudolph
\nopagebreak
\begin{center}
\parbox{15.5cm}{\sl\samepage
Institut f\"ur Experimentalphysik, Universit\"at Innsbruck, A-6020
Innsbruck, Austria$^{18}$}
\end{center}\end{sloppypar}
\vspace{2mm}
\begin{sloppypar}
\noindent
E.~Bouhova-Thacker,
C.K.~Bowdery,
D.P.~Clarke,
G.~Ellis,
A.J.~Finch,
F.~Foster,
G.~Hughes,
R.W.L.~Jones,
M.R.~Pearson,
N.A.~Robertson,
M.~Smizanska
\nopagebreak
\begin{center}
\parbox{15.5cm}{\sl\samepage
Department of Physics, University of Lancaster, Lancaster LA1 4YB,
United Kingdom$^{10}$}
\end{center}\end{sloppypar}
\vspace{2mm}
\begin{sloppypar}
\noindent
O.~van~der~Aa,
C.~Delaere,$^{28}$
G.Leibenguth,$^{31}$
V.~Lemaitre$^{29}$
\nopagebreak
\begin{center}
\parbox{15.5cm}{\sl\samepage
Institut de Physique Nucl\'eaire, D\'epartement de Physique, Universit\'e Catholique de Louvain, 1348 Louvain-la-Neuve, Belgium}
\end{center}\end{sloppypar}
\vspace{2mm}
\begin{sloppypar}
\noindent
U.~Blumenschein,
F.~H\"olldorfer,
K.~Jakobs,
F.~Kayser,
K.~Kleinknecht,
A.-S.~M\"uller,
B.~Renk,
H.-G.~Sander,
S.~Schmeling,
H.~Wachsmuth,
C.~Zeitnitz,
T.~Ziegler
\nopagebreak
\begin{center}
\parbox{15.5cm}{\sl\samepage
Institut f\"ur Physik, Universit\"at Mainz, D-55099 Mainz, Germany$^{16}$}
\end{center}\end{sloppypar}
\vspace{2mm}
\begin{sloppypar}
\noindent
A.~Bonissent,
P.~Coyle,
C.~Curtil,
A.~Ealet,
D.~Fouchez,
P.~Payre,
A.~Tilquin
\nopagebreak
\begin{center}
\parbox{15.5cm}{\sl\samepage
Centre de Physique des Particules de Marseille, Univ M\'editerran\'ee,
IN$^{2}$P$^{3}$-CNRS, F-13288 Marseille, France}
\end{center}\end{sloppypar}
\vspace{2mm}
\begin{sloppypar}
\noindent
F.~Ragusa
\nopagebreak
\begin{center}
\parbox{15.5cm}{\sl\samepage
Dipartimento di Fisica, Universit\`a di Milano e INFN Sezione di
Milano, I-20133 Milano, Italy.}
\end{center}\end{sloppypar}
\vspace{2mm}
\begin{sloppypar}
\noindent
A.~David,
H.~Dietl,
G.~Ganis,$^{27}$
K.~H\"uttmann,
G.~L\"utjens,
W.~M\"anner,
\mbox{H.-G.~Moser},
R.~Settles,
M.~Villegas,
G.~Wolf
\nopagebreak
\begin{center}
\parbox{15.5cm}{\sl\samepage
Max-Planck-Institut f\"ur Physik, Werner-Heisenberg-Institut,
D-80805 M\"unchen, Germany\footnotemark[16]}
\end{center}\end{sloppypar}
\vspace{2mm}
\begin{sloppypar}
\noindent
J.~Boucrot,
O.~Callot,
M.~Davier,
L.~Duflot,
\mbox{J.-F.~Grivaz},
Ph.~Heusse,
A.~Jacholkowska,$^{6}$
L.~Serin,
\mbox{J.-J.~Veillet}
\nopagebreak
\begin{center}
\parbox{15.5cm}{\sl\samepage
Laboratoire de l'Acc\'el\'erateur Lin\'eaire, Universit\'e de Paris-Sud,
IN$^{2}$P$^{3}$-CNRS, F-91898 Orsay Cedex, France}
\end{center}\end{sloppypar}
\vspace{2mm}
\begin{sloppypar}
\noindent
%\samepage
P.~Azzurri, 
G.~Bagliesi,
T.~Boccali,
L.~Fo\`a,
A.~Giammanco,
A.~Giassi,
F.~Ligabue,
A.~Messineo,
F.~Palla,
G.~Sanguinetti,
A.~Sciab\`a,
P.~Spagnolo
R.~Tenchini
A.~Venturi
P.G.~Verdini
\samepage
\begin{center}
\parbox{15.5cm}{\sl\samepage
Dipartimento di Fisica dell'Universit\`a, INFN Sezione di Pisa,
e Scuola Normale Superiore, I-56010 Pisa, Italy}
\end{center}\end{sloppypar}
\vspace{2mm}
\begin{sloppypar}
\noindent
O.~Awunor,
G.A.~Blair,
G.~Cowan,
A.~Garcia-Bellido,
M.G.~Green,
L.T.~Jones,
T.~Medcalf,
A.~Misiejuk,
J.A.~Strong,
P.~Teixeira-Dias
\nopagebreak
\begin{center}
\parbox{15.5cm}{\sl\samepage
Department of Physics, Royal Holloway \& Bedford New College,
University of London, Egham, Surrey TW20 OEX, United Kingdom$^{10}$}
\end{center}\end{sloppypar}
\vspace{2mm}
\begin{sloppypar}
\noindent
R.W.~Clifft,
T.R.~Edgecock,
P.R.~Norton,
I.R.~Tomalin,
J.J.~Ward
\nopagebreak
\begin{center}
\parbox{15.5cm}{\sl\samepage
Particle Physics Dept., Rutherford Appleton Laboratory,
Chilton, Didcot, Oxon OX11 OQX, United Kingdom$^{10}$}
\end{center}\end{sloppypar}
\vspace{2mm}
%\pagebreak
\begin{sloppypar}
\noindent
\mbox{B.~Bloch-Devaux},
D.~Boumediene,
P.~Colas,
B.~Fabbro,
E.~Lan\c{c}on,
\mbox{M.-C.~Lemaire},
E.~Locci,
P.~Perez,
J.~Rander,
B.~Tuchming,
B.~Vallage
\nopagebreak
\begin{center}
\parbox{15.5cm}{\sl\samepage
CEA, DAPNIA/Service de Physique des Particules,
CE-Saclay, F-91191 Gif-sur-Yvette Cedex, France$^{17}$}
\end{center}\end{sloppypar}
%\nopagebreak
\vspace{2mm}
\begin{sloppypar}
\noindent
A.M.~Litke,
G.~Taylor
\nopagebreak
\begin{center}
\parbox{15.5cm}{\sl\samepage
Institute for Particle Physics, University of California at
Santa Cruz, Santa Cruz, CA 95064, USA$^{22}$}
\end{center}\end{sloppypar}
\pagebreak
\vspace{2mm}
\begin{sloppypar}
\noindent
C.N.~Booth,
S.~Cartwright,
F.~Combley,$^{25}$
P.N.~Hodgson,
M.~Lehto,
L.F.~Thompson
\nopagebreak
\begin{center}
\parbox{15.5cm}{\sl\samepage
Department of Physics, University of Sheffield, Sheffield S3 7RH,
United Kingdom$^{10}$}
\end{center}\end{sloppypar}
\vspace{2mm}
\begin{sloppypar}
\noindent
A.~B\"ohrer,
S.~Brandt,
C.~Grupen,
J.~Hess,
A.~Ngac,
G.~Prange
\nopagebreak
\begin{center}
\parbox{15.5cm}{\sl\samepage
Fachbereich Physik, Universit\"at Siegen, D-57068 Siegen, Germany$^{16}$}
\end{center}\end{sloppypar}
\vspace{2mm}
\begin{sloppypar}
\noindent
C.~Borean,
G.~Giannini
\nopagebreak
\begin{center}
\parbox{15.5cm}{\sl\samepage
Dipartimento di Fisica, Universit\`a di Trieste e INFN Sezione di Trieste,
I-34127 Trieste, Italy}
\end{center}\end{sloppypar}
\vspace{2mm}
\begin{sloppypar}
\noindent
H.~He,
J.~Putz,
J.~Rothberg
\nopagebreak
\begin{center}
\parbox{15.5cm}{\sl\samepage
Experimental Elementary Particle Physics, University of Washington, Seattle,
WA 98195 U.S.A.}
\end{center}\end{sloppypar}
\vspace{2mm}
\begin{sloppypar}
\noindent
S.R.~Armstrong,
K.~Berkelman,
K.~Cranmer,
D.P.S.~Ferguson,
Y.~Gao,$^{13}$
S.~Gonz\'{a}lez,
O.J.~Hayes,
H.~Hu,
S.~Jin,
J.~Kile,
P.A.~McNamara III,
J.~Nielsen,
Y.B.~Pan,
\mbox{J.H.~von~Wimmersperg-Toeller}, 
W.~Wiedenmann,
J.~Wu,
Sau~Lan~Wu,
X.~Wu,
G.~Zobernig
\nopagebreak
\begin{center}
\parbox{15.5cm}{\sl\samepage
Department of Physics, University of Wisconsin, Madison, WI 53706,
USA$^{11}$}
\end{center}\end{sloppypar}
\vspace{2mm}
\begin{sloppypar}
\noindent
G.~Dissertori
\nopagebreak
\begin{center}
\parbox{15.5cm}{\sl\samepage
Institute for Particle Physics, ETH H\"onggerberg, 8093 Z\"urich,
Switzerland.}
\end{center}\end{sloppypar}
}
\footnotetext[1]{Also at CERN, 1211 Geneva 23, Switzerland.}
\footnotetext[2]{Now at Fermilab, PO Box 500, MS 352, Batavia, IL 60510, USA}
\footnotetext[3]{Also at Dipartimento di Fisica di Catania and INFN Sezione di
 Catania, 95129 Catania, Italy.}
\footnotetext[4]{Now at University of Florida, Department of Physics, Gainesville, Florida 32611-8440, USA}
\footnotetext[5]{Also Istituto di Cosmo-Geofisica del C.N.R., Torino,
Italy.}
\footnotetext[6]{Also at Groupe d'Astroparticules de Montpellier, Universit\'{e} de Montpellier II, 34095, Montpellier, France.}
\footnotetext[7]{Supported by CICYT, Spain.}
\footnotetext[8]{Supported by the National Science Foundation of China.}
\footnotetext[9]{Supported by the Danish Natural Science Research Council.}
\footnotetext[10]{Supported by the UK Particle Physics and Astronomy Research
Council.}
\footnotetext[11]{Supported by the US Department of Energy, grant
DE-FG0295-ER40896.}
\footnotetext[12]{Now at Departement de Physique Corpusculaire, Universit\'e de
Gen\`eve, 1211 Gen\`eve 4, Switzerland.}
\footnotetext[13]{Also at Department of Physics, Tsinghua University, Beijing, The People's Republic of China.}
\footnotetext[14]{Supported by the Leverhulme Trust.}
\footnotetext[15]{Permanent address: Universitat de Barcelona, 08208 Barcelona,
Spain.}
\footnotetext[16]{Supported by Bundesministerium f\"ur Bildung
und Forschung, Germany.}
\footnotetext[17]{Supported by the Direction des Sciences de la
Mati\`ere, C.E.A.}
\footnotetext[18]{Supported by the Austrian Ministry for Science and Transport.}
\footnotetext[19]{Now at SAP AG, 69185 Walldorf, Germany}
\footnotetext[20]{Now at Groupe d' Astroparticules de Montpellier, Universit\'e de Montpellier II, 34095 Montpellier, France.}
\footnotetext[21]{Now at BNP Paribas, 60325 Frankfurt am Mainz, Germany}
\footnotetext[22]{Supported by the US Department of Energy,
grant DE-FG03-92ER40689.}
\footnotetext[23]{Now at Institut Inter-universitaire des hautes Energies (IIHE), CP 230, Universit\'{e} Libre de Bruxelles, 1050 Bruxelles, Belgique}
\footnotetext[24]{Also at Dipartimento di Fisica e Tecnologie Relative, Universit\`a di Palermo, Palermo, Italy.}
\footnotetext[25]{Deceased.}
\footnotetext[26]{Now at SLAC, Stanford, CA 94309, U.S.A}
\footnotetext[27]{Now at INFN Sezione di Roma II, Dipartimento di Fisica, Universit\`a di Roma Tor Vergata, 00133 Roma, Italy.}
\footnotetext[28]{Research Fellow of the Belgium FNRS}
\footnotetext[29]{Research Associate of the Belgium FNRS} 
\footnotetext[30]{Now at Liverpool University, Liverpool L69 7ZE, United Kingdom} 
\footnotetext[31]{Supported by the Federal Office for Scientific, Technical and Cultural Affairs through
the Interuniversity Attraction Pole P5/27}   
\setlength{\parskip}{\saveparskip}
\setlength{\textheight}{\savetextheight}
\setlength{\topmargin}{\savetopmargin}
\setlength{\textwidth}{\savetextwidth}
\setlength{\oddsidemargin}{\saveoddsidemargin}
\setlength{\topsep}{\savetopsep}
\normalsize
\newpage
\pagestyle{plain}
\setcounter{page}{1}

\parskip 1.7mm
\section{Introduction}

A search for squark production in \epem\ collisions is relevant at 
LEP energies because the stop quark and, to a lesser extent, the 
sbottom quark, could well be the lightest supersymmetric partner of 
all standard model fermions~\cite{drees}. 

Searches for squarks have already been performed by 
ALEPH~\cite{squark,Barate:2000qf} 
in the framework of the MSSM, the minimal supersymmetric extension of the 
standard model~\cite{susy}, with R-parity conservation and the 
assumption that the lightest supersymmetric particle (LSP) is the lightest 
neutralino or the sneutrino. These searches yielded an absolute lower limit 
on the stop mass of 63\,\gevcc\ at the 95\% confidence level. 

This absolute lower limit does not apply if the LSP is either the gluino or 
the squark itself. Supersymmetric models in which the gluino is 
the LSP are reviewed in Ref.~\cite{gunion}. Squarks as LSP's are 
cosmologically disfavoured because of their nonzero electric 
charge~\cite{ellis}, but could be sufficiently stable to behave like the LSP 
in the LEP detectors. Scenarios in which the gluino is the next-to-lightest 
supersymmetric particle (NLSP)  and decays 
into a b quark and a long-lived sbottom (with mass 2 to 5\,\gevcc) have 
been proposed to explain excesses in \bbbar\ production at hadron 
colliders~\cite{berger}.

In this paper, it is assumed that the LSP is either the gluino or a squark, 
in the context of the MSSM with R-parity conservation. Under this
assumption, the NLSP would decay with a 100\% branching fraction into 
the LSP, {\it e.g.}, $\st\ \to {\rm c}\glu$, $\sbt\ \to {\rm b}\glu$ or
$\glu \to \sbt \bar{\rm b}$. (Stop decays into ${\rm t}\glu$ and gluino 
decays into $\st \bar{\rm t}$ can 
only happen for stop and gluino masses larger than the top quark mass, which 
is beyond the reach of LEP.) This coloured LSP is stable and hadronizes
into stable, colourless, charged and
neutral bound states. These bound states ({\it e.g.}, \glu g, \glu\qqbar,
\glu qqq, \sqqbar, \sq qq) are called R-hadrons, in which ``R''
refers to the fact that they carry one unit of R-parity~\cite{Farrar:1978xj}.

An important consequence is that the missing energy signature
might not automatically be present, in contrast to other searches for 
supersymmetric particles in which the LSP is a weakly-interacting particle,
{\it i.e.}, a neutralino or a sneutrino. However, when the LSP is 
sufficiently heavy, the maximum energy available 
for R-hadronic interactions turns out to be quite small. Indeed, the 
centre-of-mass energy $E_{\rm scat}$ of the R-hadronic interaction with 
ordinary matter, made of nucleons of mass $m_{\rm N}$, is very close 
to the mass $m_{\rm R}$ of the R-hadron, almost independently of the R-hadron 
energy $E_{\rm R}$: 
$E_{\rm scat}^2 = m_{\rm R}^2 + m_{\rm N}^2 + 2 m_{\rm N}E_{\rm R}$.
It leaves little energy ($E_{\rm scat}- m_{\rm R} -m_{\rm N}$) 
for each R-hadronic interaction in the calorimeters. For example, for 
$m_{\rm R} = 50\,\gevcc$ and $E_{\rm R}=90$\,GeV, this energy amounts 
to 800\,MeV, to be compared to $\sim 12$\,GeV for a pion of the same energy.

Because of this reduced hadronic interaction, a large missing energy signal 
is expected from heavy neutral R-hadrons. Heavy charged R-hadrons are 
expected to interact mostly electromagnetically like heavy muons. Searches 
for missing energy and for stable heavy charged particles are therefore well 
suited to look for stable squarks and gluinos. 

In view of covering all possible configurations in the plane (\mglu, \msq) 
with a squark or a gluino LSP (or, equivalently, with a long-lived squark or 
gluino), the processes investigated in this paper are
\begin{enumerate}
\item the \epemto\ \qqbar\glu\glu\ process, with a gluon splitting into 
a pair of stable gluinos;

\item the \epemto\ \qqbar\sqsqbar\ process, with a gluon splitting into 
a pair of stable squarks;

\item the pair production of stable squarks, \epemto\ \ststbar\ 
and \sbsbbar;

\item the stop pair production with decays into
stable gluinos, $\epemto \ \ststbar\ \to {\rm c}\glu \bar {\rm c}\glu$.
\end{enumerate}
The first two processes were searched for using the data collected by 
ALEPH at LEP\,1, at centre-of-mass energies around the Z resonance. 
These data correspond to about 4.5 million hadronic Z decays. The data 
collected at LEP\,2 were used to analyse the last 
two processes. The integrated luminosities and centre-of-mass energies 
of these data are indicated in Table~\ref{tab:lum}.

\begin{table}[htbp]
\caption{\footnotesize Integrated luminosities, centre-of-mass energy ranges 
and mean centre-of-mass energy values for the data collected during 
the years 1997-2000.
\label{tab:lum}}
\begin{center}
\begin{tabular}{|c|c|c|c|}
\hline
\hline
Year&Luminosity [\invpb]&Energy range [GeV]& 
$\langle\sqrt{s}\rangle$ [GeV]\\
\hline
2000&9.4&207-209&208.0\\
 &122.6&206-207&206.6\\
 &75.3&204-206&205.2\\
\hline
1999&42.0 & - & 201.6\\
 &86.2 & - & 199.5\\
 &79.8 & - & 195.5\\
 &28.9 & - & 191.6\\
\hline
1998&173.6&-&188.6\\
\hline
1997&56.8&-&182.7\\
\hline
\hline
\end{tabular}   
\end{center}
\end{table}

This paper is organized as follows. The detector is briefly described in 
Section~2. The simulation of the signal final states is discussed in 
detail in Section~3, and a brief account of the simulation of the 
standard model backgrounds is given. The 
\epemto\ \qqbar\glu\glu\ and 
\qqbar\sq\sq\ searches at LEP\,1 are presented in Sections~4 and~5, followed 
by the search for stable squark pair production at LEP\,2 in 
Section~6, and by the search for decaying stop quarks in Section~7. 
The result of the combination of all these analyses is given in 
Section~8. 

\section{The ALEPH detector}

A detailed description of the ALEPH detector and of its performance can 
be found in Refs.~\cite{aleph,perf}. Only a summary is given here.

Charged particles are detected in the central part, consisting of a 
precision silicon vertex detector, a cylindrical drift chamber and a 
large time projection chamber, which measure altogether up to 31 space 
points along the charged particle trajectories. The time projection 
chamber also provides 359 measurements (338 from wires and 21 from pads) of 
the specific energy loss by ionization ${\mathrm d}E/{\mathrm d}x$. 
A 1.5\,T axial magnetic 
field is provided by a superconducting solenoidal coil. Charged particle 
transverse momenta are reconstructed with a 1/$p_{T}$ resolution of 
$6\times 10^{-4}  \bigoplus 5 \times 10^{-3}/p_{T}$ ($p_T$ in \gevc).
In the following, {\it good tracks} are defined as charged particle
tracks reconstructed with at least four hits in the time projection
chamber, originating from within a cylinder of length 20\,cm
and radius 2\,cm coaxial with the beam and centred at the nominal
collision point, and with a polar angle with respect to the beam
such that $\vert \cos\theta \vert < 0.95$.

Electrons and photons are identified by the characteristic longitudinal
and transverse developments of the associated showers in the electromagnetic 
calorimeter, a 22 radiation length thick sandwich of lead planes and 
proportional wire chambers with fine read-out segmentation. The
relative energy resolution achieved is $0.18/\sqrt{E}$ ($E$ in GeV)
for isolated electrons and photons.

Muons are identified by their characteristic penetration pattern in the 
hadron calorimeter, a 1.5 m thick yoke interleaved with 23 layers of 
streamer tubes, together with two surrounding double-layers of muon chambers. 
In association with the electromagnetic calorimeter, the hadron calorimeter 
also provides a measurement of the hadronic energy with a relative resolution 
of $0.85/\sqrt{E}$ ($E$ in GeV).

The total visible energy, and therefore the missing energy, is measured with 
an energy-flow reconstruction algorithm which combines all the above 
measurements, supplemented by the energy detected down to 34\,mrad 
from the beam axis (24\,mrad at LEP\,1) by two additional 
electro\-magnetic calorimeters, used for the luminosity determination. 
The relative resolution on the total visible energy is $0.60/\sqrt{E}$ 
($E$ in GeV) for high-multiplicity final states. This algorithm also provides 
a list of reconstructed objects, classified as charged particles, photons 
and neutral hadrons, called {\it energy-flow particles} in the 
following, and used to determine the event characteristics for the 
selections presented in this paper.

%Finally, jets originating from b quarks are identified with a lifetime 
%b-tagging algorithm~\cite{aleph_btag}, which takes advantage of the 
%three-dimensional impact parameter resolution of charged particle 
%tracks. For tracks with two space points in the silicon vertex 
%detector~\cite{aleph_vdet} {\it i.e.}, with $\vert \cos\theta \vert < 0.85$)
%in the second phase of LEP, this resolution can be parametrized as 
%$(34+70/p)(1+1.6\cos^4\theta)~\mu$m ($p$~in~\gevc). The measured track 
%impact parameters and their resolution are then used to determine the 
%probability for each charged particle track to originate from the 
%main interaction point. A combined probability that all tracks originate 
%from the primary vertex, $P_{\rm evt}$, is then determined to estimate 
%the b content of each event. Typically, a requirement that $P_{\rm evt}$ 
%be smaller than 0.10 rejects 90\% of the light quark flavours, and 
%retains 90\% of events with two energetic b quarks.

\section{Monte Carlo Simulation}

\subsection{Signal simulation}

\subsubsection{Production processes}

\begin{figure}[htbp]
\begin{picture}(160,95)
\put(13,-5){\epsfxsize135mm\epsfbox{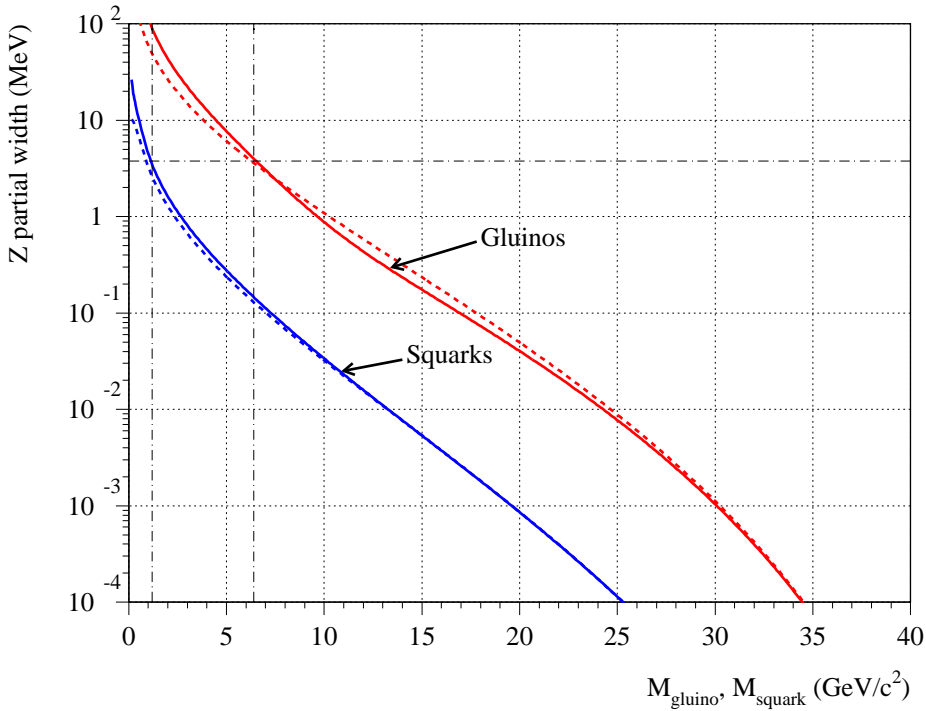}}
\end{picture}
\caption[ ]
{\protect\footnotesize The $\Z \to \qqbar\glu\glu$ and 
$\Z \to \qqbar\sqsqbar$ partial widths as a function of the squark 
and the gluino mass, computed at the leading order (dashed curves) and 
resummed at the next-to-leading order (full curves). Also shown are 
the corresponding 95\%\,C.L. lower limits on the squark and gluino 
masses when the contribution to the Z width exceeds 3.9\,MeV (Ref.~\cite{janot}
and Section~4).
\label{fig:janot}}
\end{figure}

The simulation of squark pair production at LEP\,2 energies 
was performed with the {\tt PYTHIA} event generator~\cite{Sjostrand:2000wi}. 
For the simulation of gluon 
splitting into a gluino pair at the Z resonance, the program mentioned
in Ref.~\cite{gunion} was used, modified to include initial-state radiation
as described in Ref.~\cite{remt1} and to generate events as in 
Ref.~\cite{verdier}. The QCD leading-order production cross section 
for \epemto\ \qqbar\glu\glu\ was determined numerically with this program, 
and cross-checked with the analytical predictions of 
Refs.~\cite{jezabek,seymour}, after proper quark to gluino colour-factor 
modification. These analytical formulae also allowed the 
\epemto\ \qqbar\sqsqbar\ cross
section to be determined, by appropriately modifying the quark to a squark 
phase space factor. The calculation of the resummed QCD next-to-leading-order 
cross section for gluon splitting into a gluino or a squark pair 
was performed following the prescriptions of Ref.~\cite{seymour}, with 
the running of the strong coupling constant corrected for the presence of 
a light gluino or a light squark~\cite{running}. The resulting Z partial 
widths are displayed in Fig.~\ref{fig:janot} as a function of the gluino 
and the squark mass.

\subsubsection{R-hadron formation}

The hadronization of the (s)partons generated as described above 
was done with {\tt PYTHIA}. Because squark and gluino hadronization is 
not available by default, {\tt PYTHIA} had to be extended with a few 
dedicated routines~\cite{private}. The following issues are addressed 
therein.

\paragraph{Stable gluino and squark hadronization}

Stable gluinos and squarks are allowed to radiate gluons~\cite{Norrbin:2000uu} 
as any other coloured particles. The fragmentation is handled with a Peterson 
function~\cite{peter} with a parameter extrapolated~\cite{para}
from its value for b quarks: 
\begin{equation}
\frac{\epsilon_{\sq,\glu}}{\epsilon_b}=
\frac{m_{\rm b}^2}{m_{\sq,\glu}^2}.
\end{equation}
However, a  gluino is attached to two, rather than one, string pieces.
In this case, the fragmentation is applied to each of the two pieces, 
thereby giving a larger energy loss than for a (s)quark.

Both R-meson and R-baryon production are possible. All R-hadrons 
made of a given squark or gluino are assumed to be equal in mass, with 
an electric charge of $-1$,~0~or~1, determined 
by the quark and squark content. The relative fractions of charged and 
neutral R-hadrons are then obtained by simple statistics, and are typically 
half and half, except for gluino R-hadrons, for which the additional 
possibility to form a gluino-gluon bound state enhances the fraction of 
neutral R-hadrons by an unknown amount.

In the processes studied in this paper, R-hadrons are produced in 
pairs, so that double neutral ($\R^0\R^0$), mixed ($\R^{0}\R^{\pm}$) or 
double charged  ($\R^{\pm}\R^{\pm}, \R^{\pm}\R^{\mp}$) final states are 
possible. There is no significant correlation between the electric 
charges of the two gluino R-hadrons. Charge conservation is ensured by 
the presence of additional fragmentation particles in the final state.

\paragraph{Stop-hadron decay} 

The stop decay to c\glu\ proceeds through loop diagrams or through 
unlikely FCNC tree-level couplings~\cite{Hikasa:1987db}, 
which leads to a lifetime larger than the typical hadronization time. 
The stop quark therefore hadronizes into a stop R-hadron before decaying. 
The decay is described in the framework of the spectator 
model~\cite{Altarelli:1982kh}, in which the bound state quarks act as 
spectators to the decaying stop.

\subsubsection{R-hadronic interaction in the calorimeters}

A major issue of the simulation is the treatment of the interaction of
R-hadrons in the detector. Extensive studies have been done on this
subject~\cite{gunion,Mafi:1999dg}. The simple approach chosen here
to simulate R-hadronic interactions, in analogy with 
Ref.~\cite{Barate:2000qf}, is to treat R-hadrons as heavy pions in the 
{\tt GHEISHA} package (in which the hadron mass is properly
accounted for), with the modification that the low energy 
pion-nucleon resonances are removed.

\subsubsection{Free parameters}

Several unknown parameters, listed below, are needed to fully specify the 
R-hadron phenomenology, and may be the source of sizeable systematic 
uncertainties.

\begin{itemize}

\item The probability $P_{\glu{\rm g}}$ to form a gluon-gluino bound state. 
This quantity is essentially unknown and can vary between 0 and 1. 
For $P_{\glu{\rm g}} = 0$, the fractions of double neutral, mixed and
double charged final states are approximately 25$\%$, 50$\%$, and 25$\%$, 
respectively. In the %unlikely 
configuration in which $P_{\glu{\rm g}} = 1$ (unlikely for gluino 
masses in excess of 25\,\gevcc~\cite{gunion,Mafi:1999dg}), only purely 
neutral final states are produced. 

\item The effective spectator mass $M^{\rm eff}_{\rm spec}$ of squark and 
gluino R-hadrons. It roughly corresponds to the difference between the 
constituent mass and the current algebra mass. The value obtained from 
hadron mass spectroscopy is $M^{\rm eff}_{\rm spec}~=~0.5^{+0.5}_{-0.2}\,\gevcc$~\cite{spectator}.

\item The gluon constituent mass $M^{\rm c}_{\rm g}$. Its value is 
estimated from glueball searches~\cite{glueball} to be $M^{\rm c}_{\rm g} = 
0.7_{-0.4}^{+0.3}\,\gevcc$, to be compared to the quark 
constituent mass, typically twice smaller.

\item The total cross section $\sigma_{\rm RN}$ of R-hadron-nucleon 
scattering. It is assumed to be equal to the cross section 
$\sigma_{\pi\rm N}$ of pion-nucleon scattering because
{\it (i)} at high momentum, the interaction cross section of any hadron 
on a nucleon is proportional to the sum of the individual valence parton 
cross sections, and {\it (ii)} heavy partons (with a mass above a couple 
of~\gevcc) do no interact significantly. The sum therefore runs only 
over the standard partons, in which each quark accounts for one unit 
of cross section, and each colour octet gluon for 9/4 units. For gluinos, 
the R-meson (\glu\qqbar) hadronic cross section is thus equal to that of 
a pion, the \glu{g} bound state cross section is $9/4/(1+1) = 9/8$ larger, 
and the R-baryon (\glu qqq) cross section is $(1+1+1)/(1+1) = 3/2$ larger. 
For squark-R-mesons (\sqqbar), the same argument yields a ratio of 1/2 while, 
for squark-R-baryons (\sq qq), the ratio is equal to unity. 
Altogether, it is therefore reasonable to use an average R-hadronic 
interaction cross section identical to that of pions, with an uncertainty 
of $\pm 50\%$.

\end{itemize}

\subsubsection{Simulated signal samples}

In order to design the signal selection criteria, hundreds of simulated 
signal samples of 1000 to 4000 events each were generated for each of the four
processes mentioned in Section~1. 
Gluon splitting to gluino or squark pairs was simulated for gluino/squark 
masses between 
0 and 40\,\gevcc, with a typical step of 2\,\gevcc. For squark pair production,
samples with squark and gluino masses ranging from 0 to 90\,\gevcc\ 
with a typical 5\,\gevcc\ step were generated, with two values of the 
squark mixing angle: {\it (i)} $0^\circ$ and {\it (ii)} the value for which 
the squark coupling to the Z vanishes, {\it i.e.}, $56^\circ$ for stops and 
$68^\circ$ for sbottoms.

\subsection{Background simulation}

Simulated samples for all relevant standard model background processes were
generated both at LEP\,1 and LEP\,2 energies. Bhabha scattering was simulated 
with the {\tt BHWIDE} generator~\cite{Jadach:1997nk}. The {\tt KORALZ} 
package~\cite{Jadach:1994yv} was used for the other difermion processes. 
Two-photon interactions were simulated with 
{\tt PHOT02}~\cite{Vermaseren:1980xg}. The $\W^+\W^-$ production
was simulated with {\tt KORALW}~\cite{Jadach:1998gi}, the production of 
$\W {\rm e} \nu$ with {\tt GRC4F}~\cite{Fujimoto:1997wj} and the 
$\Z$ee, $\Z\Z$ and $\Z\nnbar$ final states with 
{\tt PYTHIA}~\cite{Sjostrand:1994yb}. 

%\noindent
%All background and signal samples 
%were processed through the full detector simulation.

\section{Search for 
{\mbox{\boldmath \epemto\ \qqbar\glu\glu}} with LEP\,1 data}

At LEP\,1, events resulting from gluon radiation off a \qqbar\ pair, with
subsequent gluon splitting into two stable gluinos, are expected to show as 
a pair of acoplanar jets, accompanied by two stable R-hadrons. 

When the gluino mass is small, the \epemto\ \qqbar\glu\glu\ production 
cross section is large enough to sizeably contribute to the Z hadronic 
width. The accurate electroweak measurements at LEP and SLC allow a 
model-independent upper limit of 3.9\,MeV to be set on the Z width 
for purely hadronic final states~\cite{janot}. All gluino masses below 
6.3\,\gevcc\ are therefore excluded at the 95\% confidence level~\cite{janot},
as can be seen in Fig.~\ref{fig:janot}, irrespective of the gluino decay 
and hadronization mechanisms. For larger masses, the final state topology, 
and therefore the search strategy, depends on the electric charges of 
the R-hadrons. 

\subsection{Search for two neutral R-hadrons}

\begin{figure}[htbp]
\begin{picture}(160,87)
\put(15,-5){\epsfxsize130mm\epsfbox{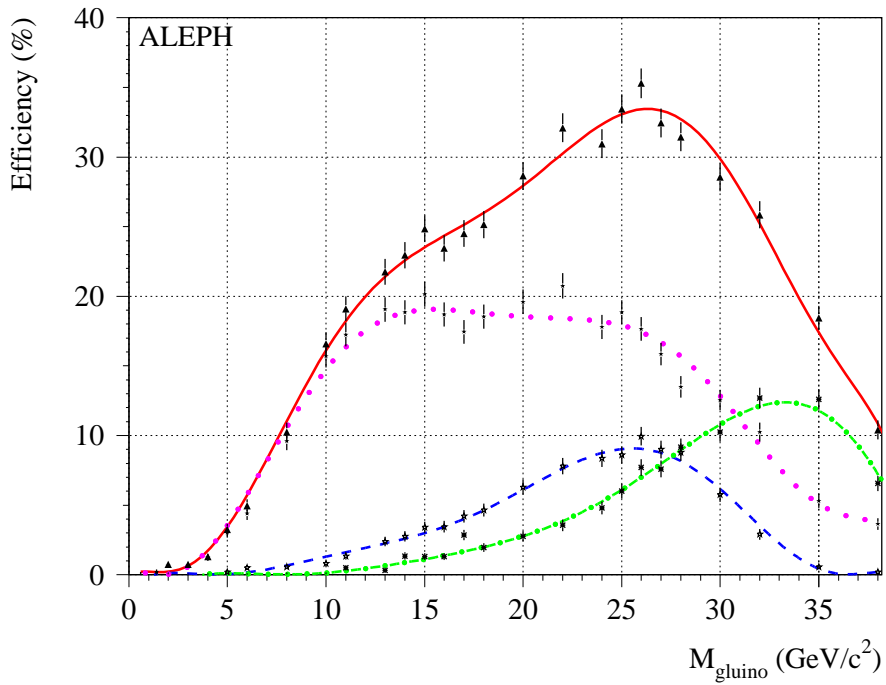}}
\end{picture}
\caption[ ]
{\protect\footnotesize The \qqbar\glu\glu\ selection efficiency 
as a function of the gluino mass, for the H\nnbar\ selection (dashed
curve), the additional efficiency of the large-mass selection 
(dot-dashed curve), the additional efficiency of the small-mass
selection (dotted curve), and the sum of the three (full curve). The markers 
with error bars indicate the efficiencies obtained with each of the 
simulated samples, and the curves are obtained with a polynomial fit 
through these points.
\label{fig:efficgg}}
\end{figure}

When the two R-hadrons are neutral, the acoplanar jet pair is 
accompanied by missing energy arising from the specific interaction 
of massive neutral R-hadrons with the detector. The acoplanar jet 
selection developed for the H\nnbar\ search~\cite{lep1} may therefore 
be used. Its efficiency reaches $\sim 9\%$ for $\mglu\ = 25\,\gevcc$,
as is visible from the dashed curve in Fig.~\ref{fig:efficgg}.
This analysis was optimized for the case of a heavy Higgs boson, leading to 
non-violently acoplanar hadronic jets and a moderate, but isolated, missing 
energy, which makes it inadequate for gluino masses above 30 and below 
15\,\gevcc. 

\subsubsection{Large gluino masses}

For gluino masses above 30\,\gevcc, the missing energy and the 
jet acoplanarity become so large that simpler and, in this 
configuration, more efficient selection criteria can be designed with the 
same topological variables as in the H\nnbar\ search. Such selection criteria 
were optimized for a search for the stop quark~\cite{dudu}, and are 
summarized below.

Only events with at least four good tracks are considered. Most of the 
\qqbar\ background is rejected by the requirement that the visible mass
be smaller than 50\%$\sqrt{s}$. Events with a small angle $\theta_{\rm T}$
of the thrust axis with respect to the beam ($\cos\theta_{\rm T} > 0.9$)
are eliminated, and so are events with energy detected within $12^\circ$
from the beam axis, to avoid events in which energy escapes down the beam 
pipe. The bulk of two-photon interactions is rejected by requiring that
the total momentum transverse to the beam exceed 5\%$\sqrt{s}$. 

\parskip 1.2mm
For events with at least eight  good tracks (mostly \qqbar\ events), 
the acollinearity and the acoplanarity angles 
are required to be smaller than $135^\circ$ and $150^\circ$, respectively.
Three-jet events with one energetic neutrino from heavy-quark semi-leptonic 
decays are rejected by requiring that the total momentum transverse to 
the beam remain smaller than 30\,\gevc. To reject three-jet events with two 
such neutrinos, the aplanarity ({\it i.e.}, the sum of the three 
jet-jet angles when the event is forced to form three jets) has to be less 
than $350^\circ$. 

In events with less than eight good tracks, the thrust axis angle with 
respect to the beam axis has to be larger than $45^\circ$. For 
monojet events, {\it i.e.}, events in which one hemisphere (with respect
to a plane perpendicular to the thrust axis) is empty, it is 
further required that the visible mass be in excess of 5\,\gevcc\ and 
the thrust smaller than 0.95. These cuts eliminate the 
two-photon interactions and the $\epemto\ \tau^+\tau^-$ background.
For non-monojet events, each hemisphere must contain at least two good 
tracks and the acoplanarity angle has to be less than $150^\circ$, which 
rejects the remaining $\tau^+\tau^-$ background. 

The additional efficiency brought by this large-mass selection 
exceeds 10\% for \mglu\ above 30\,\gevcc, as can be seen from the 
dot-dashed curve in Fig.~\ref{fig:efficgg}.

\subsubsection{Small gluino masses}

For gluino masses below 15\,\gevcc, the R-hadrons tend to interact more
in the calorimeters, and a third jet, mostly formed with neutral hadronic
energy, develops around the still large missing energy. Requirements like
missing energy isolation or substantial aplanarity therefore become
inefficient, and must be replaced by more selective criteria aimed at 
this specific three-jet topology.

Only hadronic events (more than five good tracks carrying more than 10\% of 
the centre-of-mass energy) with a visible mass smaller than 65\,\gevcc\ are 
considered here. As in the H\nnbar\ selection, the acollinearity and the 
acoplanarity angles are required to be less than $165^\circ$ and $175^\circ$, 
respectively. Similarly, events with less than 75\% of the visible energy 
above $30^\circ$ from the beam axis are eliminated, and so are the events with 
any activity below $12^\circ$. Two-photon interactions are rejected 
by requiring that the missing transverse momentum exceed 
$5\%\sqrt{s}$ for events with a visible mass below 20\,\gevcc, and 
that the missing momentum point more than $35^\circ$ away from 
the beam axis. The latter
cut also rejects \qqbar\ events with an undetected, energetic, 
initial-state-radiation photon.

The remaining events are then forced to form three jets with 
the Durham algorithm. To ensure a three-jet topology, the 
values where the transitions from three to two jets ($y_{23}$) 
and to four jets ($y_{34}$) occur are required to exceed 0.01 and to be 
less than 0.04, respectively.

\begin{figure}[htbp]
\begin{picture}(160,160)
\put(18,78){\epsfxsize125mm\epsfbox{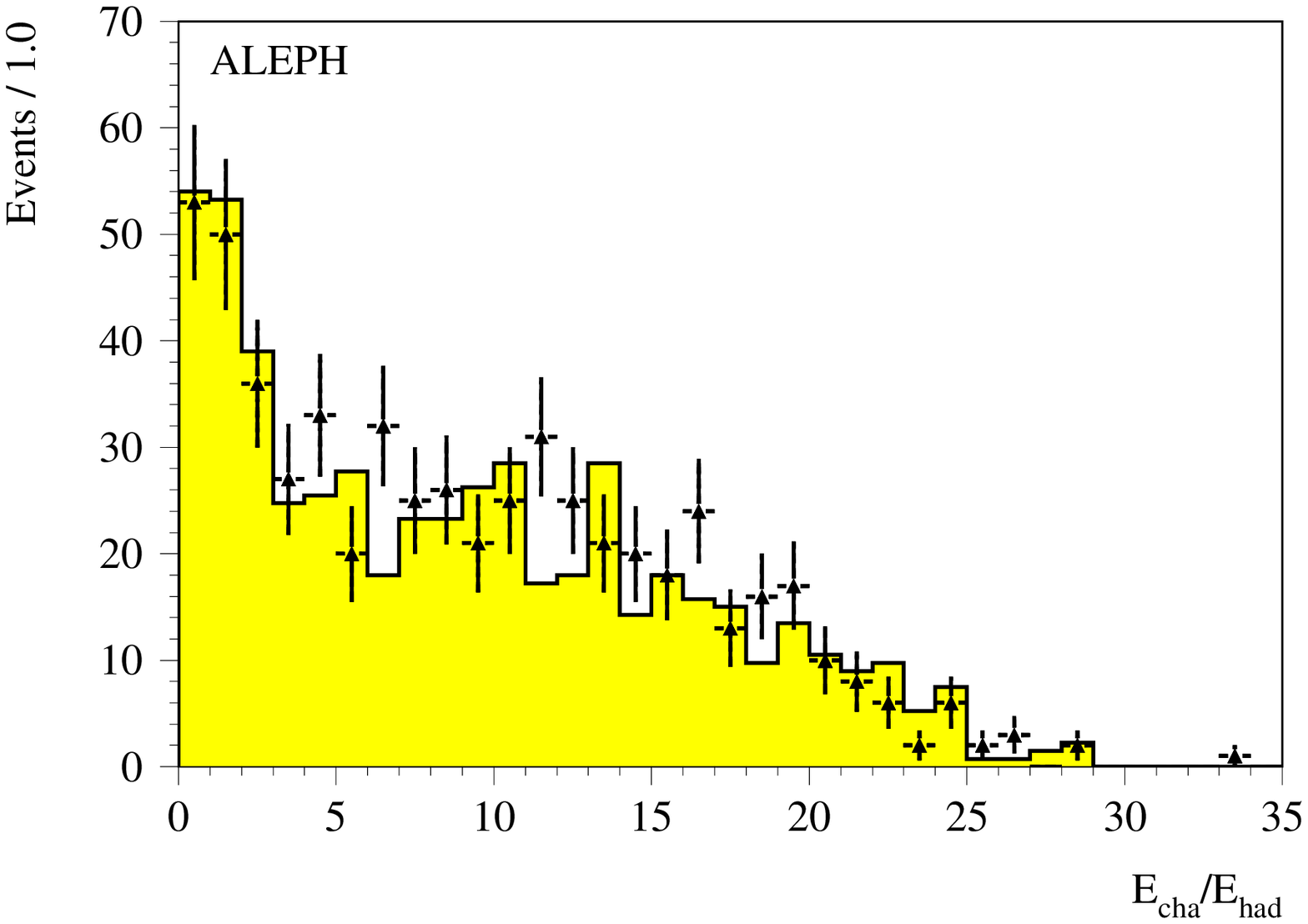}}
\put(67.55,113.75){\epsfxsize70mm\epsfbox{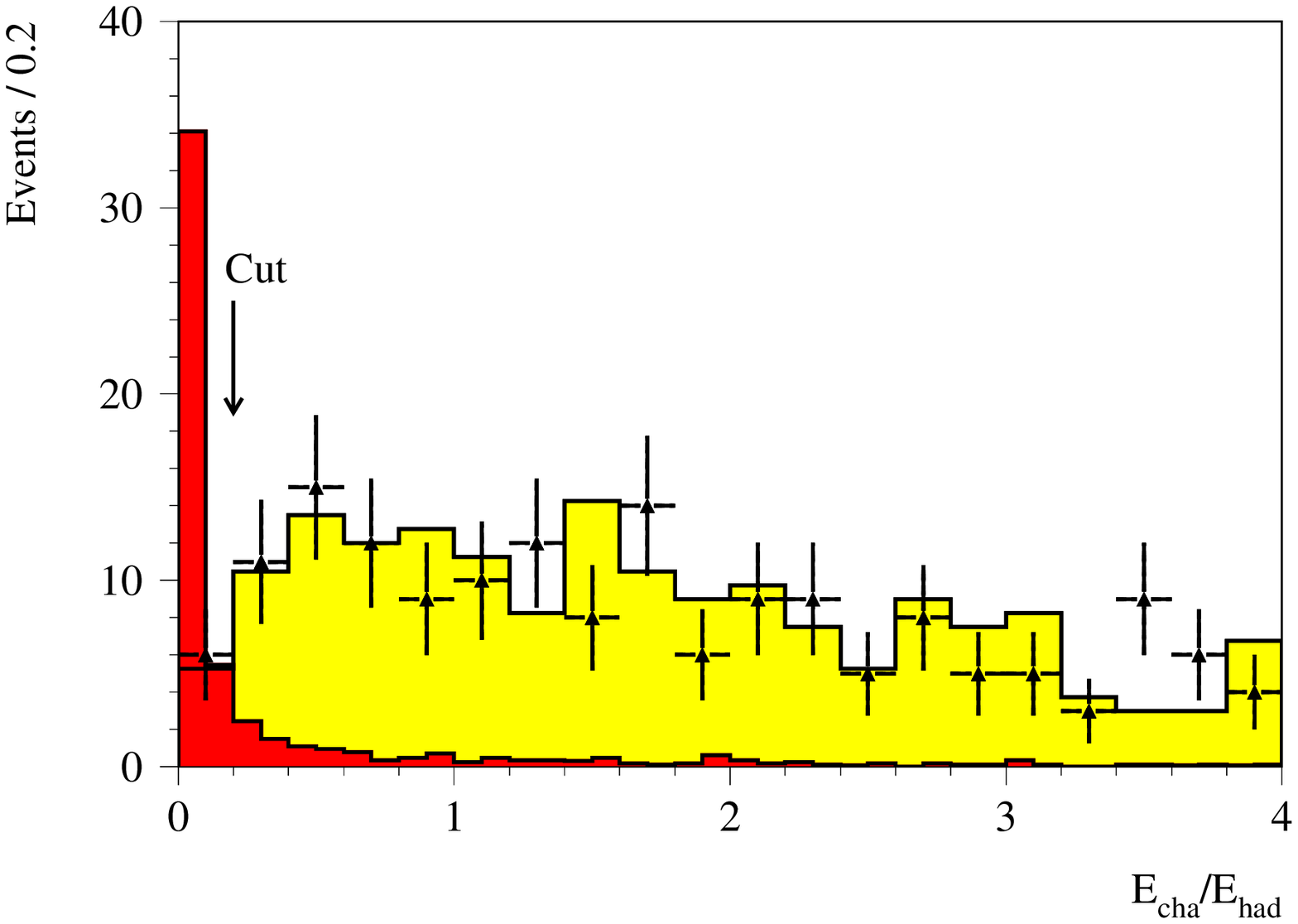}}
\put(18,-5){\epsfxsize125mm\epsfbox{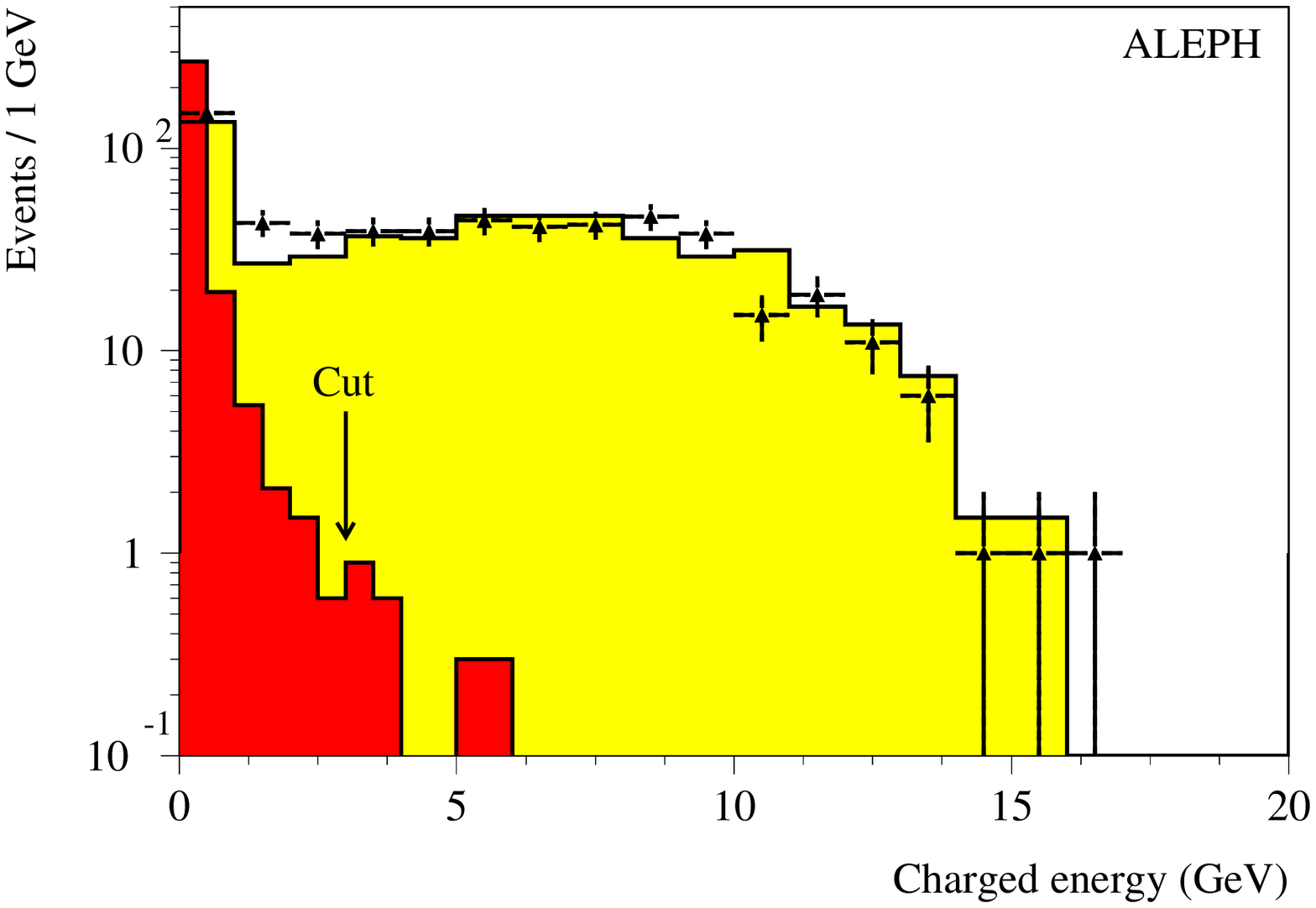}}
\put(85,153){Zoom between 0. and 4.}
\put(122,98){(a)}
\put(122,15){(b)}
\end{picture}
\caption[ ]
{\protect\footnotesize Distributions of (a) the ratio of the ``R-hadron jet'' 
charged energy to its neutral hadronic energy; and (b) the charged energy 
in a $35^\circ$ cone around the missing momentum, before the 
corresponding cuts are applied, for the data (triangles with error bars), 
the background simulation (light-shaded histogram) and the signal with 
$\mglu = 10\,\gevcc$ (dark-shaded histogram, arbitrary normalization).
\label{fig:threeg}} 
\end{figure}

To select signal-like events, the ``R-hadron jet'' candidate, {\it i.e.},
the jet with the smallest amount of charged energy, is required to contain 
mostly neutral hadrons, by imposing that its charged energy be smaller than 
20\% of its neutral hadronic energy. (The latter is computed as the energy
sum of all neutral energy-flow particles in the jet with a hadron calorimeter 
component.) Because the missing energy is also expected to point in the 
direction of the ``R-hadron jet'' candidate, it is finally required that the 
amount of charged energy in a cone of $35^\circ$ 
around the missing momentum direction be smaller than 3\,GeV. 
As can be inferred from Fig.~\ref{fig:threeg}, these last two criteria cut the 
expected background by a factor larger than a thousand, but only reduce 
the signal efficiency by a modest 30\%.
\null\vskip .1cm
%\eject

\parskip 2mm
The additional efficiency brought by this last selection is shown by the 
dotted curve in Fig.~\ref{fig:efficgg}. As anticipated, the
selection is well suited for masses around 15\,\gevcc\ and below. 
Although not specifically designed for large masses, the selection also 
contributes substantially all the way up to 40\,\gevcc. 

When the three searches are combined, the \qqbar\ro\ro\ selection efficiency 
exceeds 15\% for masses between 10 and 35\,\gevcc, and culminates at about 
35\% for $\mglu\ = 27\,\gevcc$. It decreases fast for small gluino masses 
but it does not completely vanish, even for massless gluinos. Indeed, 
when two energetic light neutral R-hadrons are produced, they are 
expected to be detected as normal neutral hadrons in the hadron calorimeter. 
Due to calorimeter resolution effects, this large neutral hadronic 
energy occasionally causes sizeable missing energy and acoplanarity. 

No events were selected in the data in either of the three selections, with 
an expected background of less than two events at 95\% confidence level.

\subsection{Search for charged R-hadrons}

If the probability $P_{\glu{\rm g}}$ to form a gluon-gluino bound state
is not 100\%, a relevant fraction of the final states contains one 
or two charged R-hadrons (up to 75\% for $P_{\glu{\rm g}}=0$). 
These heavy stable charged particles can 
be identified with their isolation from the rest of the event, as well
as their unusual energy loss by ionization in the time projection chamber.

\begin{figure}[htbp]
\begin{picture}(160,150)
\put(1,75){\epsfxsize82mm\epsfbox{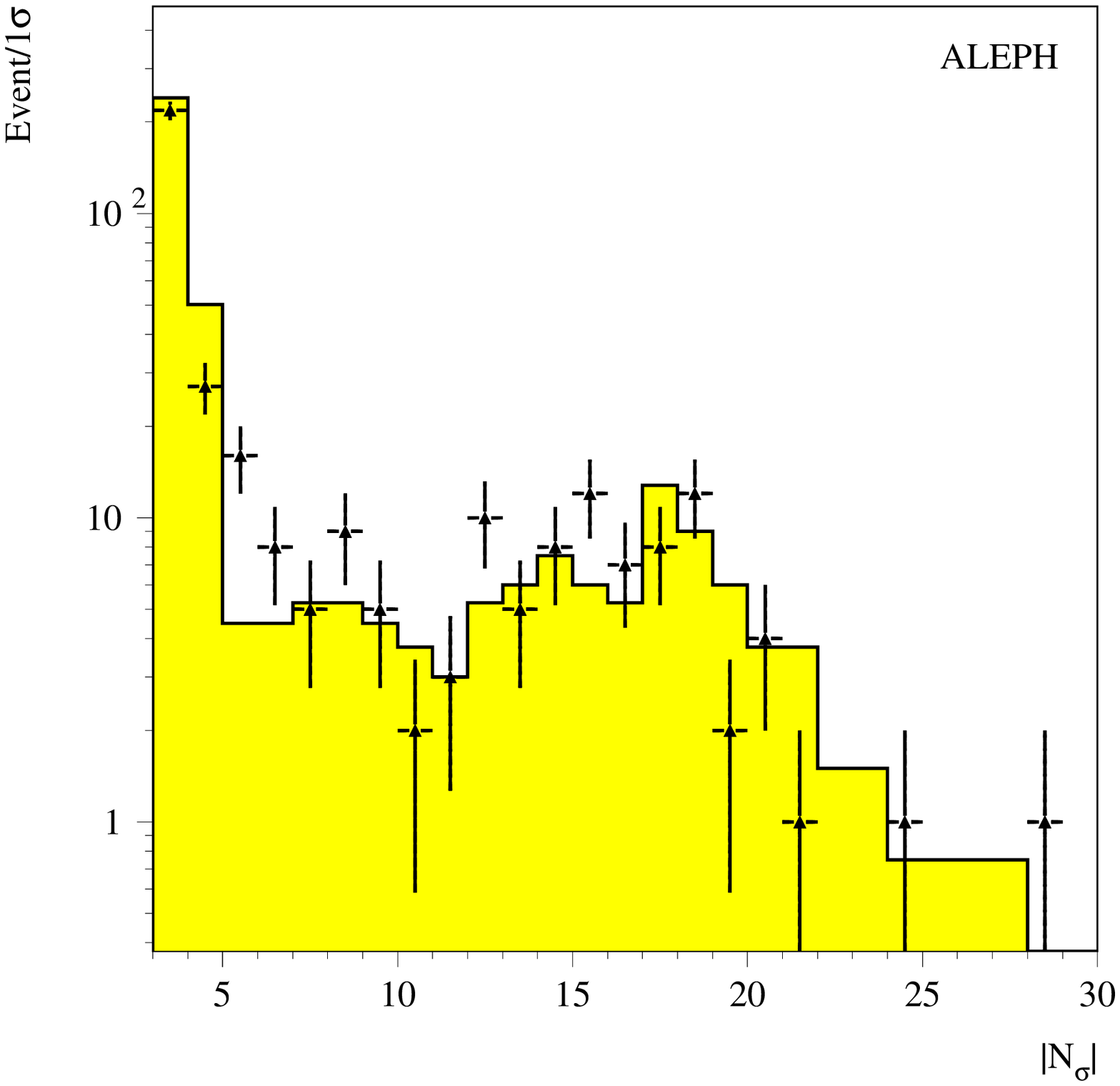}}
\put(25,141){(a)}
\put(104.5,141){(b)}
\put(25,62){(c)}
\put(104.5,62){(d)}
\put(81,75){\epsfxsize82mm\epsfbox{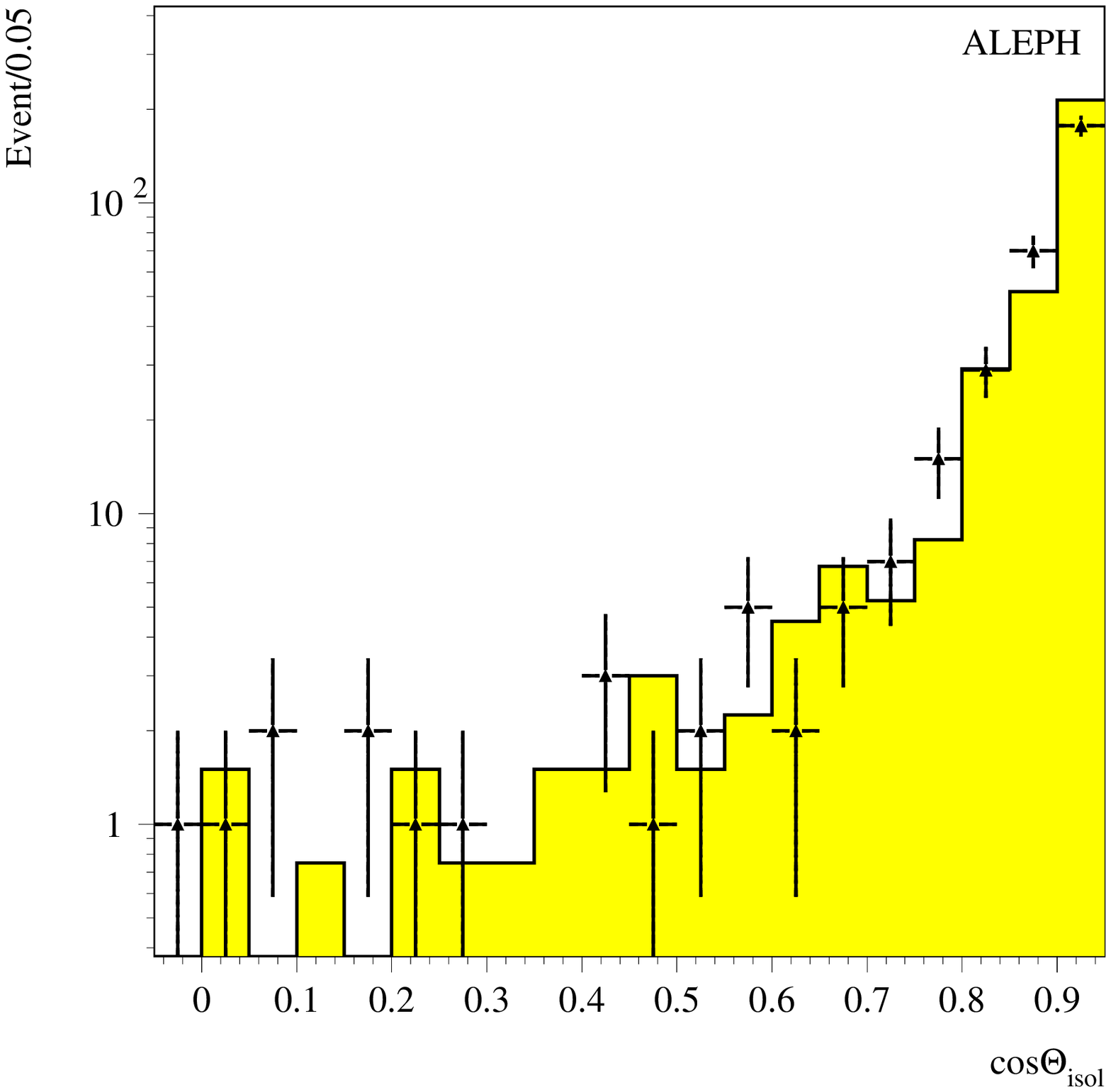}}
\put(1,-3){\epsfxsize82mm\epsfbox{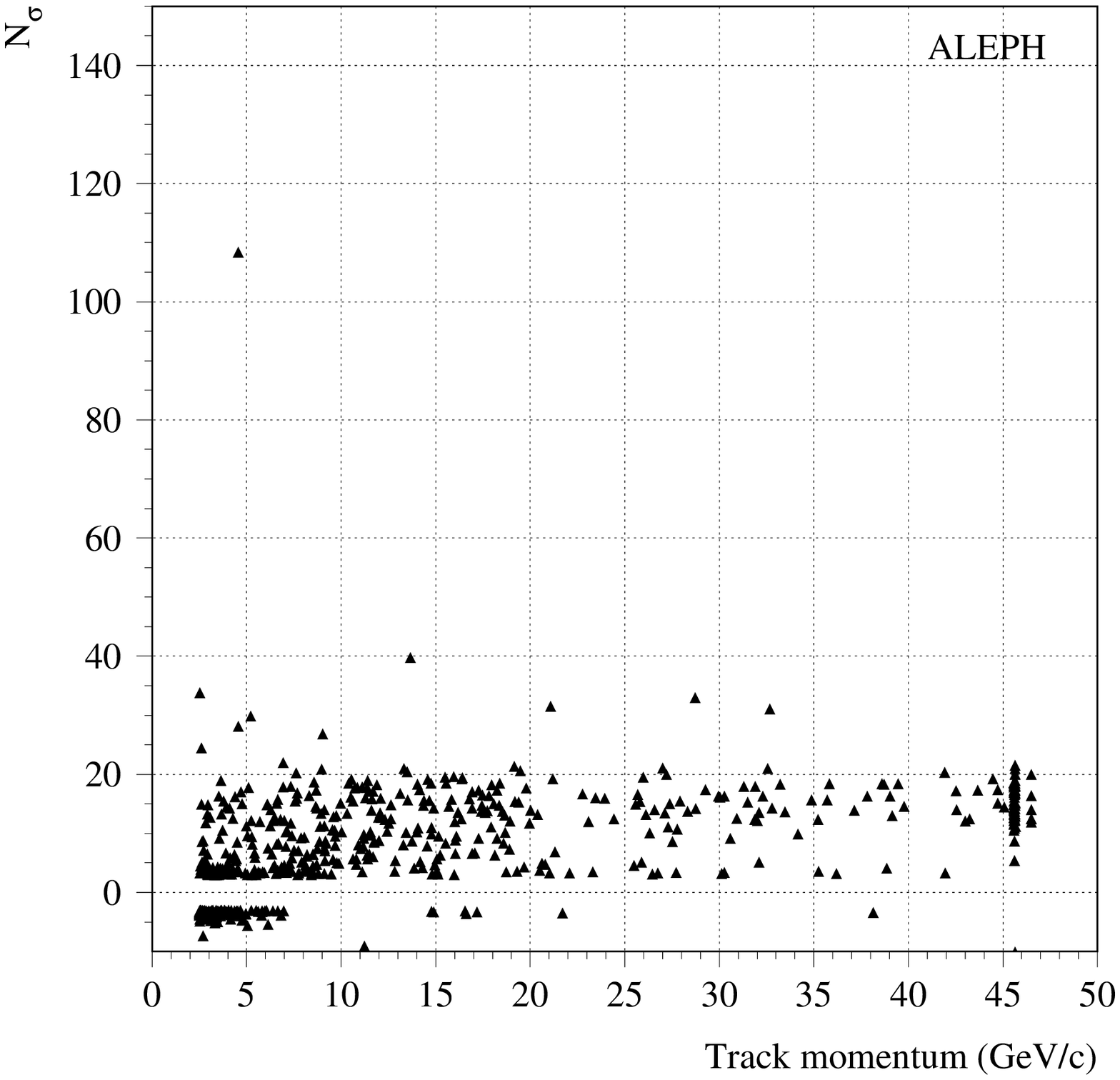}}
\put(81,-3){\epsfxsize82mm\epsfbox{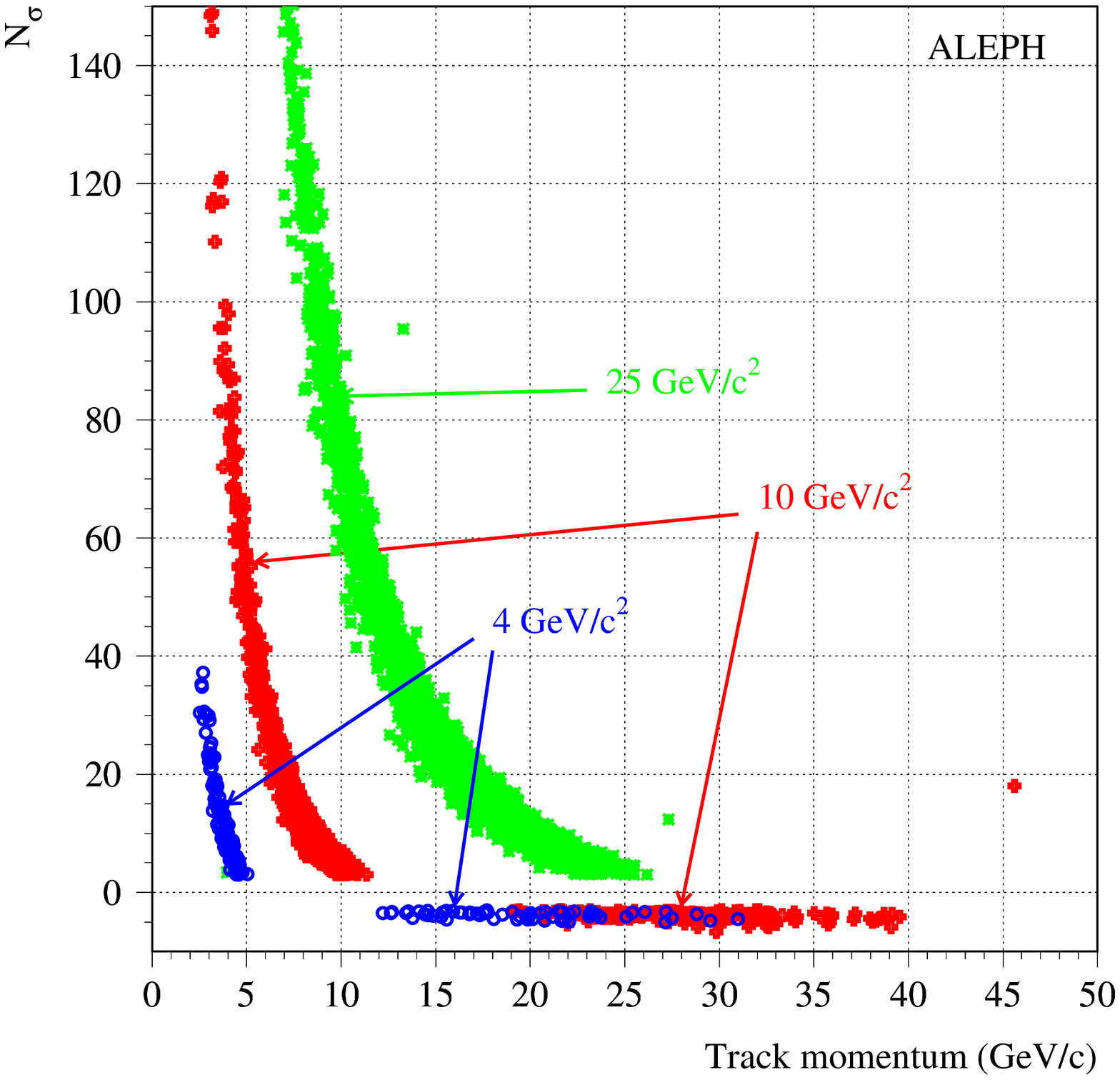}}
\end{picture}
\caption[ ]
{\protect\footnotesize Distributions of (a) $\vert N_\sigma \vert$, the 
number of standard deviations with respect to the expected dE/dx for the 
most-likely light-particle hypothesis; and (b) the cosine of the isolation 
angle, for the heavy-stable-charged-particle track candidates after the 
preselection, in the data (triangles with error bars) and the background 
simulation (histogram).
Distributions of $N_\sigma$ as a function of the charged particle 
momentum (c) in the data and (d) in the signal simulation with 
$\mglu = 4$, 10 and 25\,\gevcc.
\label{fig:dedx}} 
\end{figure}

Events with at least five good tracks are considered in this search.
Heavy stable charged particle track candidates must be reconstructed 
with at least half of the 338 wires and the 21 pads of the time projection
chamber, must have a momentum larger than 2.5\,\gevc, and must be isolated 
from the other good tracks of the event by more than $18.2^\circ$. To 
ensure a reliable momentum determination, the track-fit-$\chi^2$ probability 
is required to exceed 1\%. The energy loss by ionization is then required 
to be at least three standard deviations away from that expected for light 
stable charged particles (electrons, muons, pions, kaons or protons) of the 
same momentum. After this preselection, 
a total of 364 events with at least one heavy stable charged particle track 
candidate is selected from the data, in agreement with the $363 \pm 16$ events 
expected. The distribution of $N_\sigma$, the number of standard deviations 
with respect to the ionization expected in the most-likely light-particle 
hypothesis, is displayed in 
Fig.~\ref{fig:dedx}a, and that of the cosine of the isolation angle 
in Fig.~\ref{fig:dedx}b. 
A reasonable agreement is observed between the
data and the simulation in both variables. In particular, the dE/dx
distribution of very ionizing particles (mostly alpha particles) is well 
reproduced by the simulation.

The distributions of $N_\sigma$ as a function of the track candidate momentum
are shown in Fig.~\ref{fig:dedx}c for the data, and in Fig.~\ref{fig:dedx}d
for signal events with different gluino masses. 
Events with two heavy stable charged particle track candidates are kept as 
$\qqbar\rp\rp$ candidates. Events with only one such track, {\it i.e.}, 
$\qqbar\rp\ro$ candidates, are required to have a visible mass smaller than 
75\,\gevcc, to account for the presence of a neutral R-hadron. The track
candidate must also satisfy one of the two following tightened identification 
criteria in order to reject the remaining background: 
\begin{enumerate}
\item[{\it (i)}] either the track is isolated from the other good tracks of 
the event by more than $25.8^\circ$ and its fit-$\chi^2$ probability is in 
excess of 10\%, in which case its energy loss by ionization is required to 
be at least four standard deviations away from that expected for light 
stable charged particles of the same momentum; 
\item[{\it (ii)}] or its momentum is larger than 5\,\gevc, in which case the
energy loss by ionization is required to be {\it smaller} than expected 
for light charged particles.
\end{enumerate}

%\null\vskip .5cm
\eject
\null\vskip -0.5cm

\parskip 1mm
\subsection{Combined selection efficiency}

\begin{figure}[htbp]
\begin{picture}(160,90)
\put(15,-4){\epsfxsize130mm\epsfbox{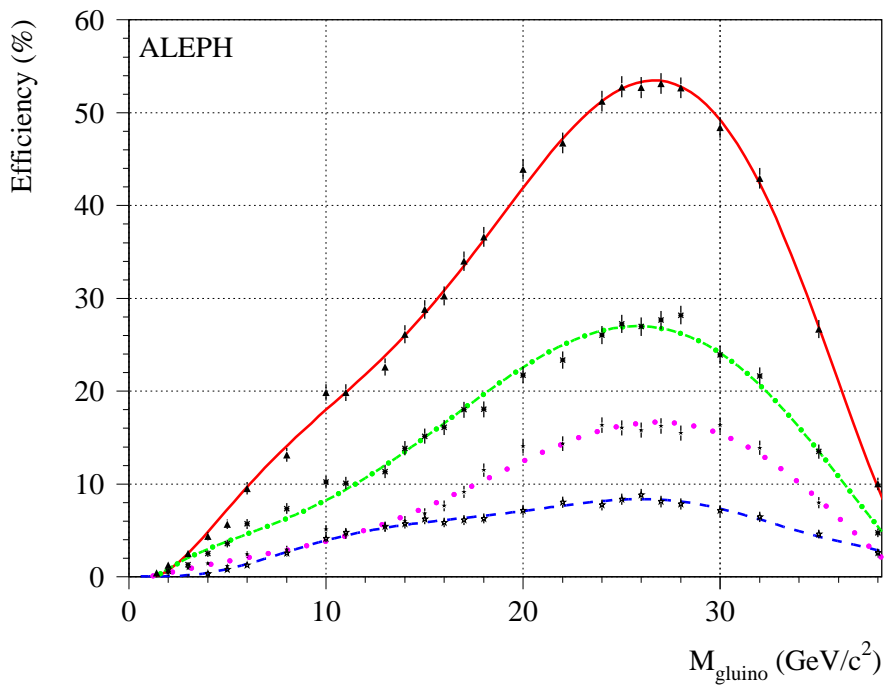}}
\end{picture}
\caption[ ]
{\protect\footnotesize The \qqbar\glu\glu\ selection efficiency 
as a function of the gluino mass, when $P_{\glu{\rm g}} = 0\%$, 
for the \qqbar\ro\ro\ final state (dashed curve), the \qqbar\rp\ro\ 
final state (dot-dashed curve), the \qqbar\rp\rp\ final state (dotted curve), 
and the sum of the three (full curve). The markers with error bars indicate 
the efficiencies obtained with each of the simulated samples, and the curves 
are obtained with a polynomial fit through these points.
\label{fig:efficqq}}
\end{figure}

The efficiencies of the combination of all (charged and neutral) R-hadron 
selections, weighted by the expected production fractions, are displayed in 
Fig.~\ref{fig:efficqq} as a function of the gluino mass for 
$P_{\glu{\rm g}} = 0\%$, in each of the three different final states, 
\qqbar\ro\ro, \qqbar\rp\ro\ and \qqbar\rp\rp. 
The total efficiency is maximal for gluino masses
in the vicinity of 27\,\gevcc, where it exceeds 50\% (to be compared
to 35\% when $P_{\glu{\rm g}} = 100\%$).
No events were selected in the data by the charged hadron selection, with 
0.7 events expected from standard model backgrounds, mostly from 
\epemto\ \qqbar.

\subsection{Systematic uncertainties}

Potentially large systematic uncertainties may arise from several 
sources and are listed below. Their relative effects on the number of signal 
events expected are given here for a gluino mass of 27\,\gevcc, but are 
similar for all masses.
\begin{itemize}
\item The hadronization is simulated with parton shower evolution, but the
parton shower parameters are mostly unknown in the presence of gluinos. 
A variation of these parameters by $\pm 100\%$ yields a $\pm 3\%$
variation of the selection efficiency.
\item The R-hadronic interaction cross section uncertainty of $\pm 50\%$
changes the selection efficiency by $\pm 9\%$ for $P_{\glu{\rm g}}=1$ and
by $\pm 3\%$ for $P_{\glu{\rm g}}=0$.
\item The gluon constituent mass uncertainty yields an efficiency 
variation of $\pm 2\%$ for $P_{\glu{\rm g}}=1$. The other gluino 
hadronization parameters of Section~3 have no sizeable effects.
\item The uncertainty on $\alpha_S(\mZ) = 0.1183 \pm 0.0020$ turns into 
a variation of $\pm 4\%$ of the production cross section.
\item The QCD next-to-leading-order corrections to the cross section 
are at the percent level for gluino masses around 25\,\gevcc, and are 
included in the present estimate. The next-order
corrections are not expected to have any visible effect.
\end{itemize}
The above uncertainties were taken into account according to 
Ref.~\cite{cousins}. 

\subsection{Gluino mass limit}

\begin{figure}[htbp]
\begin{picture}(160,87)
\put(15,-5){\epsfxsize130mm\epsfbox{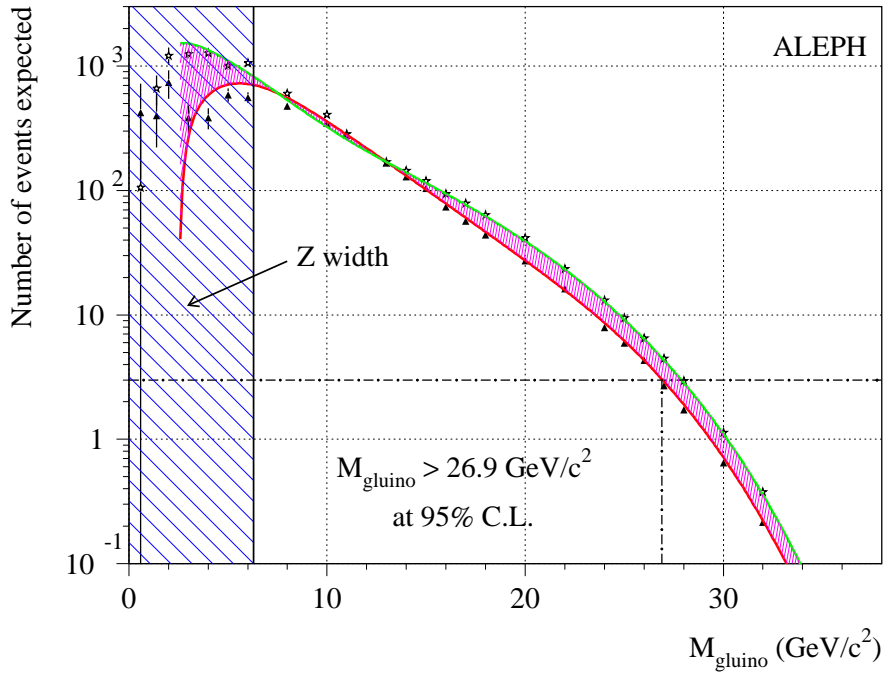}}
\end{picture}
\caption[ ]
{\protect\footnotesize The number of \qqbar\glu\glu\ events expected
to be selected in the LEP\,1 data by the neutral and charged R-hadron 
searches,  as a function of the gluino mass, for $P_{\glu{\rm g}}$ 
varying between 0 and 1 (shaded area). The markers with error bars 
(triangles for $P_{\glu{\rm g}}=1$, stars for $P_{\glu{\rm g}}=0$) show the
individual values obtained from each of the simulated signal samples. 
Also shown is the region excluded by the 
Z lineshape measurement (hatched region), and the lower limit on the 
gluino-LSP mass (dot-dashed lines).
\label{fig:expect}} 
\end{figure}

The number of 
\qqbar\glu\glu\ events expected in the LEP\,1 data is displayed in 
Fig.~\ref{fig:expect} as a function of the gluino mass, for $P_{\glu{\rm g}}$ 
varying between 0 and 1. 
Altogether, when combined to the analysis of Ref.~\cite{janot}, this search 
results in a 95\%\,C.L. absolute lower limit of 26.9\,\gevcc\ on the mass of 
a gluino LSP (Fig.~\ref{fig:expect}), which substantially improves the 
2 to 18\,\gevcc\ exclusion of Ref.~\cite{delphi} and exceeds the expectation
from the study of Ref.~\cite{gunion}.

\section{Search for {\mbox{\boldmath \qqbar\sqsqbar}} with 
LEP\,1 data}

The same selections were applied with no modification to stable squark
production through the \epemto\ \qqbar\sqsqbar\ process. 
As shown in Fig.~\ref{fig:janot}, the precise electroweak measurements allow 
all squark masses below 1.3\,\gevcc\ to be excluded with the method of 
Ref.~\cite{janot}, irrespective of the squark decay and hadronization. 

\begin{figure}[htbp]
\begin{picture}(160,90)
\put(15,-4){\epsfxsize130mm\epsfbox{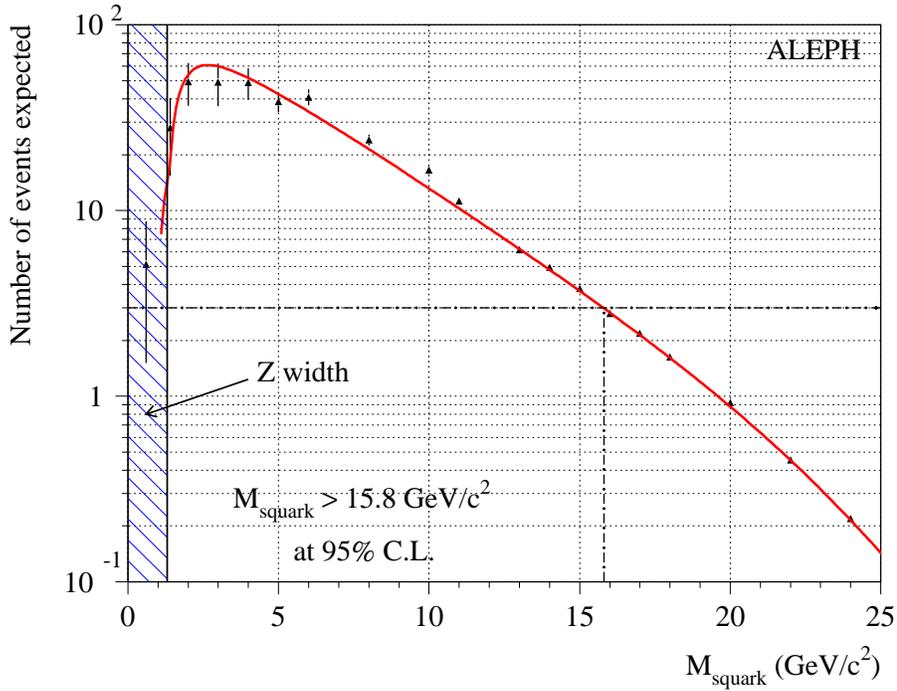}}
\end{picture}
\caption[ ]
{\protect\footnotesize The number of \qqbar\sqsqbar\ events expected
to be selected in the LEP\,1 data as a function of the squark mass 
(full curve). The triangles with error bars show the individual values 
obtained from each of the simulated
signal samples. Also shown is the region excluded by the Z lineshape 
measurement (hatched region), and the lower limit on the squark-LSP
mass (dot-dashed lines).
\label{fig:sq_exp}} 
\end{figure}

\begin{figure}[htbp]
\begin{picture}(160,90)
\put(15,-4){\epsfxsize130mm\epsfbox{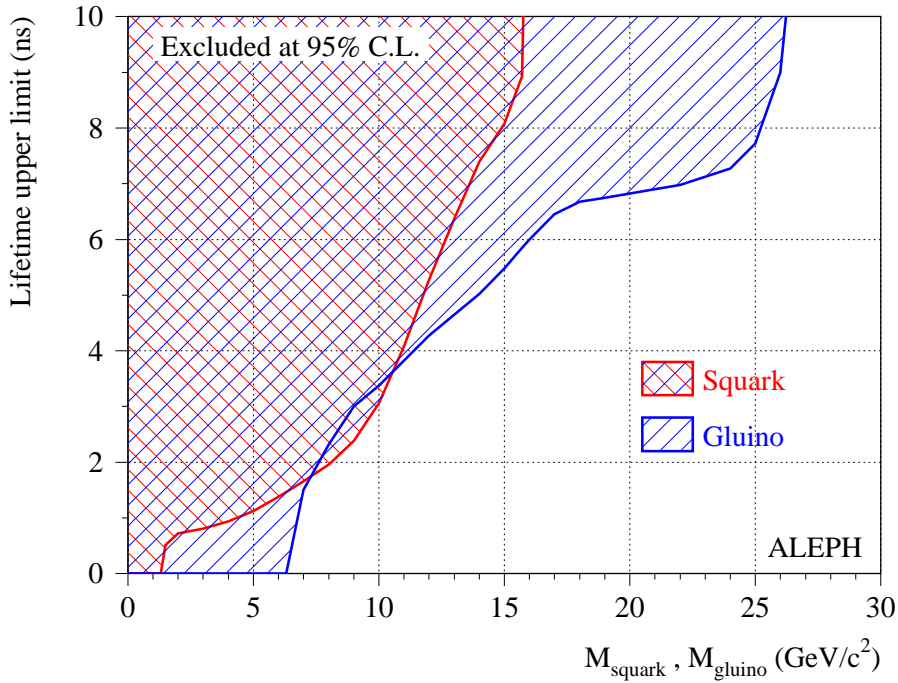}}
\end{picture}
\caption[ ]
{\protect\footnotesize The upper limit on the squark and gluino lifetimes
as a function of their mass (full curve). The hatched areas are excluded
at more than 95\% C.L. 
\label{fig:life}} 
\end{figure}

The number of events expected to be selected by the charged and neutral
R-hadron searches described above is displayed in Fig.~\ref{fig:sq_exp} 
for larger squark masses. 
This search results in a 95\%\,C.L. lower limit of 
15.7\,\gevcc\ on the mass of a stable squark. The
result was translated into upper limits on lifetimes, as 
presented in Fig.~\ref{fig:life} for squarks and gluinos.
Conservatively, the selection efficiency was assumed to vanish for 
neutral-R-hadron decays within the calorimeter volume, and for 
charged-R-hadron decays within the tracking volume.
The scenario of Ref.~\cite{berger} is excluded for sbottom lifetimes in 
excess of 1\,ns.

\parskip 1mm
\section{Search for stable squarks in 
{\mbox{\boldmath \epemto\ \sqsqbar}} at LEP\,2}

If the lightest supersymmetric particle is a squark or, more generally, if 
the lightest squark does not decay within the detector volume, its pair
production gives rise to a final state with two stable squark-R-hadrons, 
of which at least one is charged in at least 60\% of the 
cases~\cite{Barate:2000qf}. These charged R-hadrons can then be 
identified using the kinematic characteristics of squark-pair production 
and the large specific ionization ${\rm d}E/{\rm d}x$ measured with 
the time projection chamber, as described in the previous section.

\begin{figure}[htbp]
\begin{picture}(160,90)
\put(25,-30){\epsfxsize105mm\epsfbox{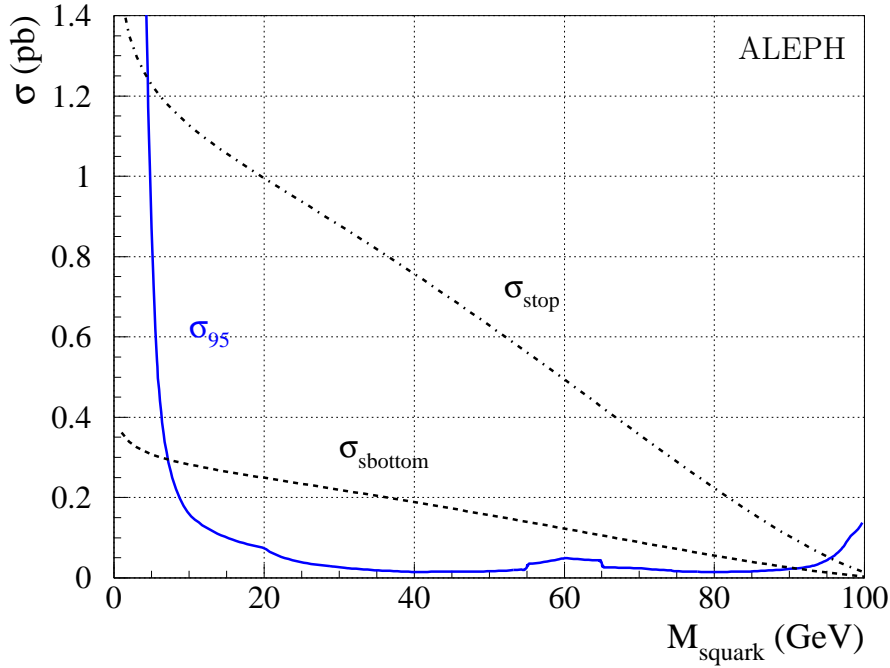}}
\put(115,80){ALEPH}
\end{picture}
\caption[ ]
{\protect\footnotesize The production cross sections for \epemto\ \ststbar\ 
and \sbsbbar\ at $\sqrt{s} = 209$\,GeV, as a function of the squark mass. 
Also shown is the observed 95\%\,C.L. upper limit on the cross section, 
derived from the stable stop search of Ref.~\cite{Barate:2000qf}.
\label{fig:sguazz}} 
\end{figure}

The stable stop search of Ref.~\cite{Barate:2000qf} was extended to lower 
stop masses so as to provide an overlap with the search of Section~5, and 
to sbottom pair production. The production cross sections for stops and 
sbottoms, and the corresponding upper limits derived from the stable stop 
search, are shown in Fig.~\ref{fig:sguazz} %as a function of the squark 
%mass, 
for a vanishing coupling to the Z.

This stable heavy charged particle search therefore allows stable stop 
masses between 5 and 95\,\gevcc\ and stable sbottom masses between 7 and 
92\,\gevcc\ to be excluded at 95\%\,C.L. When combined with the result of 
the search for \epemto\ \qqbar\sqsqbar\ with LEP\,1 data presented in the 
previous section, stable hadronizing squarks are excluded below 
95\,\gevcc\ (up-squarks) and 92\,\gevcc\ (down-squarks).

\parskip 1mm

\section{Search for  
{\mbox{\boldmath $\epemto\ \ststbar \to {\rm c}\glu\bar{\rm c}\glu$}} 
at LEP\,2}

If the gluino is the LSP and the lighter stop quark the NLSP, the 
stop then decays into a gluino and a c or a u quark. The corresponding decay 
widths~\cite{Hikasa:1987db} are such that, for all practical purposes, 
the stop quark is essentially stable if the mass difference $\Delta M$
between the stop-R-hadron and the gluino-R-hadron masses is smaller than 
the D mass, while it decays promptly if $\Delta M > m_{\rm D}$, {\it i.e.}, 
when the $\st\ \to {\rm c}\glu$ decay channel is open.
In the former case, the stable stop search of Section~6 applies here too 
with no modification. In the latter case, the final state topology is a 
pair of acoplanar jets from the \ccbar\ hadronization accompanied by 
missing energy carried away by the mass of the two gluinos. 
The selection of this final state and its results are discussed in
Section~7.1. The $\Delta M \simeq m_{\rm D}$ case, in which the stop 
lifetime can take any value, is addressed in Section 7.2.

\subsection{The case of prompt stop decays}

\subsubsection{Event selection}

The discriminant 
variables listed below were designed for generic acoplanar jet topologies:

\begin{itemize}

\item the visible mass $M_{\rm vis}$ and energy $E_{\rm vis}$, 
computed with all energy-flow particles;

\item the energy  $E_{12}$ detected within $12^\circ$ from the beam axis;

\item the total energy carried by neutral hadrons, $E_{\rm NH}$;

\item the energy of the most energetic lepton, $E_{\ell}$; 

\item the energy in a $30^\circ$ half-angle cone around the most energetic 
lepton, $E_{\ell}^{30}$;

%\item the energy of the most energetic energy-flow particle, 
%$E_{\rm lead EF}$;
%
%\item the energy of the most energetic neutral hadron, $E_{\rm lead NH}$;

\item the energy computed without the identified leptons, $E_{\rm had}$;

\item the energy measured in a $30^\circ$ azimuthal wedge around the 
direction of the missing transverse momentum, $E_{\rm Wedge}$;

\item the energy measured beyond $30^\circ$ from the beam axis, $E_{30}$;

\item the momentum transverse to the beam axis, computed with all 
energy flow particles, $p_T$, without neutral hadrons, $p_T^{\rm exNH}$ and 
with good tracks only, $p_T^{\rm ch}$;

\item the number of good tracks, $N_{\rm ch}$;

\item the acoplanarity angle $\Phi_{\rm acop}$, between the directions of 
the momenta in the two event hemispheres, projected onto 
a plane perpendicular to the beam axis. Here, the hemispheres are defined 
with respect to a plane perpendicular to the thrust axis;

\item the transverse acoplanarity angle, $\Phi_{\rm acopT}$, defined as above, 
but the hemispheres are now defined with respect to the transverse thrust 
axis, {\it i.e.}, the thrust axis of the event projected onto a plane 
perpendicular to the beam axis;

\item the cosine of the polar angle of the missing momentum vector, 
$\cos\theta_{\rm miss}$;

\item the cosine of the polar angle of the thrust axis, 
$\cos\theta_{\rm T}$;

\item the event thrust $T$; 

\item the polar angle of the scattered electron, $\theta_{\rm scat}$, computed
from the missing energy and momentum under the hypothesis that the 
event originates from a $\gamma\gamma$ interaction.
\end{itemize}

A preselection was used against two-photon events, characterized by their 
small visible energy and their boost along the beam direction. To reject 
these events, $M_{\rm vis}$ was required to exceed 15\,\gevcc, $p_T$ 
to be larger than 2.5\%$\sqrt{s}$ and $N_{\rm ch}$ to be in excess of 7. 
Moreover, the angles $\Phi_{\rm acop}$ and $\Phi_{\rm acopT}$ were both 
required to be smaller than $178.5^\circ$,  $\cos\theta_{\rm miss}$ and 
$\cos\theta_{\rm T}$ smaller than 0.85 and $E_{12}$ smaller than 
0.5\%$\sqrt{s}$. 

\begin{figure}[htbp]
\begin{picture}(160,140)
\put(3,-2){\epsfxsize155mm\epsfbox{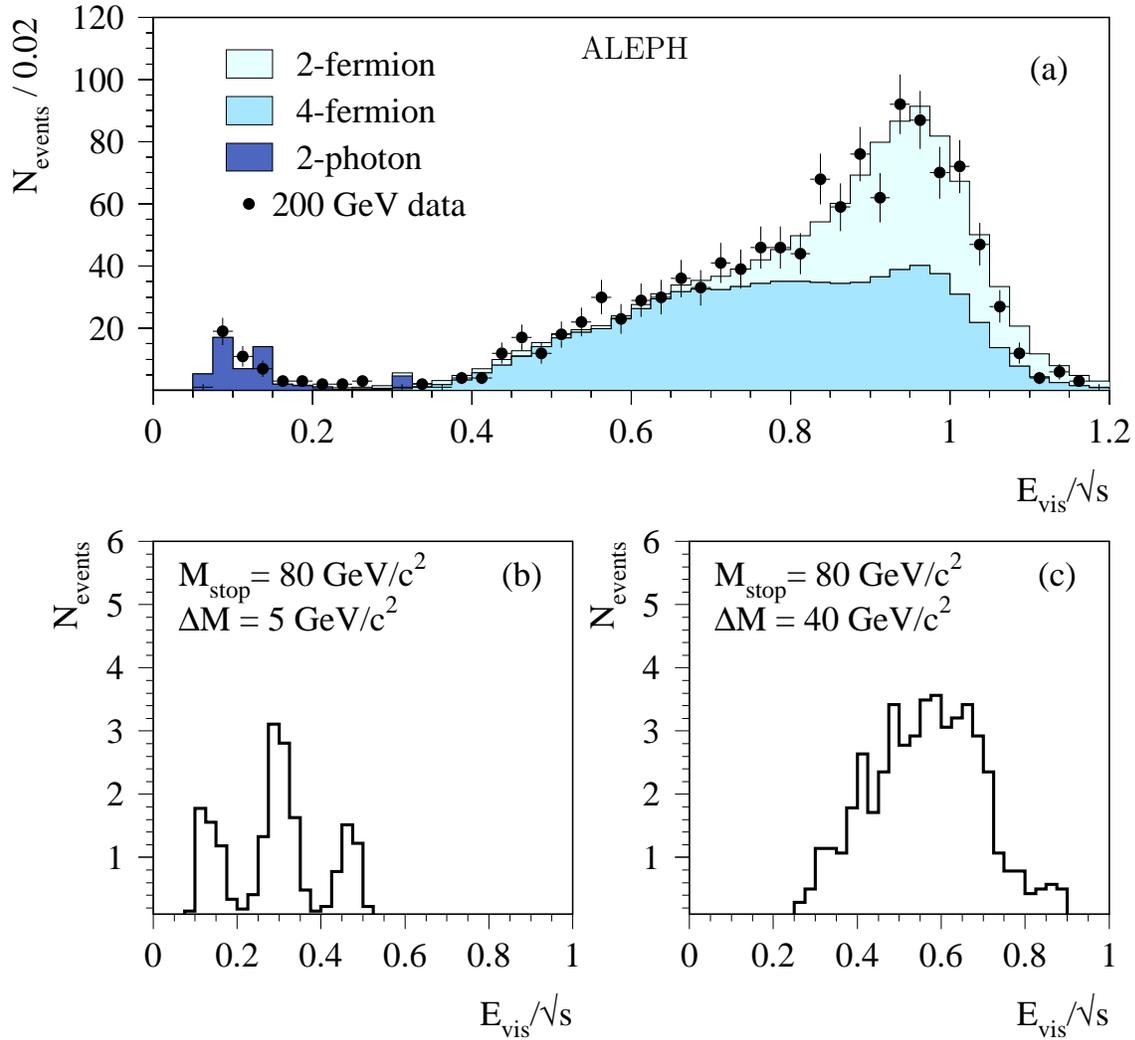}}
\put(80,132){ALEPH}
\end{picture}
\caption[ ]
{\protect\footnotesize 
Distributions of the visible energy observed in the data at $\sqrt{s} = 
200$\,GeV (dots with error bars) and expected from various background sources 
(shaded histograms), after the preselection cuts (a). The lower two plots 
show the expected distributions for the signal, with $\Delta M = 5\,\gevcc$ 
(b) and 40\,\gevcc~(c). For small $\Delta M$, the R-hadron composition is 
clearly visible: double-charged-R-hadrons events have largest visible energy, 
and double-neutral-R-hadrons events have smallest visible energy. For large
$\Delta M$, the three peaks are merged because of the broader visible energy 
distributions.
\label{fig:dist}}
\end{figure}

The distribution of the visible energy after this preselection is shown 
in Fig.~\ref{fig:dist} for the data, the background expected from different 
sources and the signal for two different values of $\Delta M$. 
For events with small $\Delta M$, only a small amount of energy is available 
for the recoiling c-quark system, which results in a small particle 
multiplicity and small visible energy. In addition, the R-hadrons tend to be 
emitted back to back, which turns into large values for thrust, acoplanarity 
and transverse acoplanarity. These characteristics are very similar to those 
of two-photon events. In contrast, events with large $\Delta M$ are 
characterized by large particle multiplicity and visible energy, and the 
background arises mainly from  four-fermion production and \qqbar\ events. 

\parskip 2mm
In fact, the distribution of most of the variables of the above list, and 
thus the relevant background sources, are strongly correlated with the 
value of $\Delta M$.~Three different selections were therefore developed 
for small, intermediate and large $\Delta M$ values.

The selection criteria follow closely those used for the squark 
searches in the case of a neutralino LSP, described in 
Ref.~\cite{Barate:1997qi}. However, the values of the most relevant cuts 
were re-optimized with the $\bar{N}_{95}$ method (that minimizes the 
expected 95\%\,C.L upper limit on the signal cross 
section~\cite{Grivaz:1992nt}) to account for the reduced missing energy, 
the larger integrated luminosity and the centre-of-mass energy increase, 
and to include the subtraction of the four-fermion and \qqbar\ backgrounds. 
The optimized cut values for the three selections are given in 
Table~\ref{tab:cuts}.

\begin{table}[htpb]
\caption{\footnotesize Selection criteria for the search for R-hadrons
from stop pair production in the acoplanar jet topology, in the three
$\Delta M$ regions.
\label{tab:cuts}}
\begin{center}
\begin{tabular}{|l|c|c|c|}
\hline\hline
Variable & Small $\Delta M$ &  Intermediate  $\Delta M$ & Large $\Delta M$ \\
\hline\hline
$M_{\rm vis}$             & $>15\,\gevcc$  & $>15\,\gevcc$   & $>15\,\gevcc$\\
$p_T/\sqrt{s}$            & $>2.5\%$       & $>4\%$          & $>7.5\%$     \\
$N_{\rm ch}$              & $>7$           & $>11$           & $>18$        \\
$E_{\rm vis}/\sqrt{s}$    & $<50\%$        & $<70\%$         & $<80\%$      \\
$E_{12}/\sqrt{s}$         & $<0.5\%$       & $<0.5\%$        & $<0.5\%$     \\
$\cos\theta_{\rm miss}$   & $>0.8$         & $>0.8$          & $>0.8$       \\
$\cos\theta_{\rm T}$      & $>0.8$         & $>0.8$          & $>0.85$      \\
$\Phi_{\rm acop}$         & $<178.5^\circ$ & $<176^\circ$    & $<176^\circ$ \\
$\Phi_{\rm acopT}$        & $>178.5^\circ$ & $<177^\circ$    & $<177^\circ$ \\
$E_{\rm Wedge}/\sqrt{s}$  & $<12.5\%$      & $<12.5\%$       & $<12.5\%$    \\
$T$                       & $<$0.96        & $<$0.94         & $<$0.92      \\
$\theta_{\rm scat}$       & $>5^\circ$     & $[6^\circ,80^\circ]$ & $[15^\circ,80^\circ]$ \\
$p_T/E_{\rm vis}$         & $-$            & $>12.5\%$       & $>12.5\%$    \\
$p_T^{\rm ch}/\sqrt{s}$   & $>1.5\%$       & $-$             & $-$          \\
$p_T^{\rm exNH}/\sqrt{s}$ & $>2\%$         & $-$             & $-$          \\
$E_{\rm had}$             & $[10\,{\rm GeV},40\%\sqrt{s}]$   &  $<55\%\sqrt{s}$ & $<75\%\sqrt{s}$  \\
$E_{\rm NH}/\sqrt{s}$     & $<10\%$         & $-$            & $-$       \\
$E_{\rm NH}/E_{\rm vis}$  & $-$             & $<30\%$        & $-$       \\
$E_{\ell}/\sqrt{s}$       & $>20\%$         & $-$            & $-$       \\
$E^{30}_{\ell}/\sqrt{s}$  & $-$             & $>1\%$         & $>1\%$    \\
$E_{30}/E_{\rm vis}$      & $-$             & $-$            & $>80\%$   \\
\hline\hline
\end{tabular}
\end{center}
\end{table}

\subsubsection{Selection efficiency}

\begin{figure}[htbp]
\begin{picture}(160,90)
\put(10,-4){\epsfxsize130mm\epsfbox{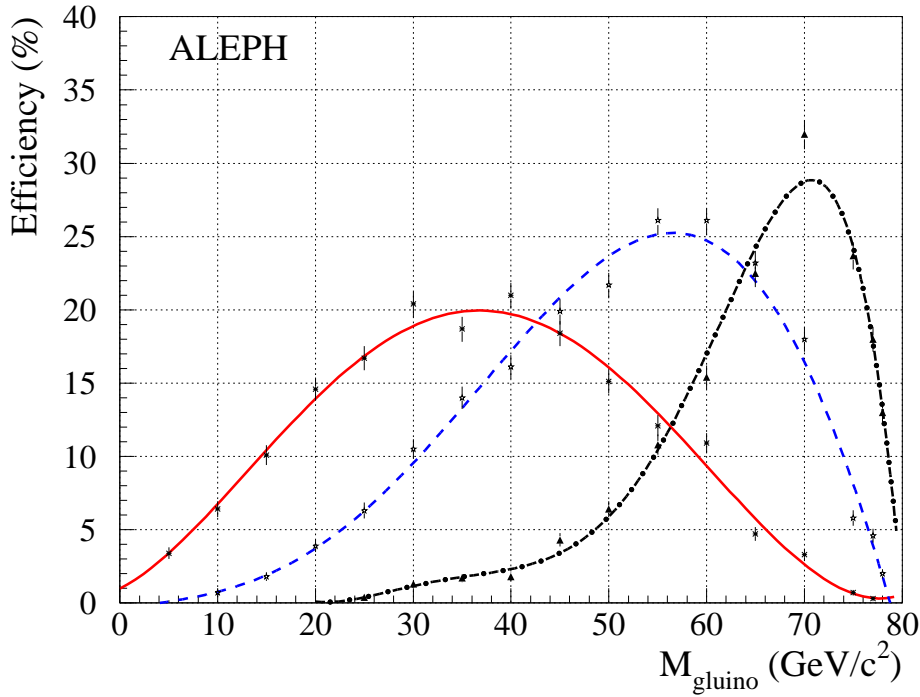}}
\end{picture}
\caption[ ]
{\protect\footnotesize 
The three parametrized selection efficiencies for $\mst = 80\,\gevcc$ (full 
curve, large $\Delta M$ selection; dashed curve, intermediate $\Delta M$ 
selection; and dot-dashed curve, small $\Delta M$ selection) as a function 
of $\mglu$. The dots with error bars show the actual efficiencies obtained
from the individual simulated samples.
\label{fig:effpar}}
\end{figure}

The selection efficiencies were obtained from simulated samples 
(Section~3.1.5) 
for over 150 different values of (\mst, \mglu), and were then parametrized as 
a function of \mst\ and \mglu\ with a polynomial function. For example, the 
efficiency of the three selections is shown in Fig.~\ref{fig:effpar} for 
$\mst = 80\,\gevcc$ as a function of the gluino mass. The $\Delta M$ intervals 
in which each of the three selections are to be used were again determined, as 
a function of the stop mass, with the $\bar{N}_{95}$ prescription. For 
example, for $\mst = 80\,\gevcc$, the switching $\Delta M$ values are 16 
and 33\,\gevcc. These switching $\Delta M$ values tend to decrease with 
the stop mass. 

\subsubsection{Systematic uncertainties}

The signal efficiencies may be affected by uncertainties in the simulation 
of the stop and gluino R-hadronization physics, uncertainties related to 
the detector response for R-hadrons and uncertainties due to the limited 
size of the simulated samples. Systematic effects from the physics assumptions 
were estimated by varying the free parameters of the simulation, as 
described in Section~3. 

\begin{itemize}
\item The gluino-gluon bound state probability was assumed to be 10\%. The 
effects of a probability variation from 0\% to 100\% depend on the stop and 
gluino masses. For light gluinos (around 30\,\gevcc), the efficiency 
increases with $P_{\glu{\rm g}}$, {\it i.e.}, with the fraction of \ro\ro\ 
final states, as is the case for the LEP\,1 neutral R-hadron selections
(Section~4.1). Indeed, such light R-hadrons substantially interact
in the calorimeters, and it is only when two neutral R-hadrons are 
present in the final state that the missing energy becomes large enough 
to make the present selection fully efficient. In contrast, for larger 
gluino masses and small mass differences, the two gluinos are 
produced approximately back to back with similar energies, and have almost 
no interaction in the calorimeters. This configuration leads to a cancellation 
in the measured missing transverse momentum in the \ro\ro\ final state. In 
this case, the efficiency decreases when $P_{\glu{\rm g}}$ increases. A 
map of the corresponding uncertainty (up to $\pm 20\%$ in extreme cases) was 
built in the (\mst, \mglu) plane.

\item The R-hadronic interaction cross section uncertainty of $\pm 50\%$
changes the selection efficiency by at most $\pm 5\%$ in the small 
$\Delta M$ region, and by a negligible amount otherwise.

\item The effective spectator mass uncertainty yields relative efficiency 
variations of $\pm 5\%$, $\pm 3\%$ and $\pm 3\%$ in the small, 
intermediate and large $\Delta M$ regions, respectively. The gluon 
constituent mass uncertainty has no visible effect.

\item The uncertainty on the Peterson fragmentation parameter 
$\epsilon_{\rm b}$ (varied here between 0.003 and 0.010) gives rise to 
an efficiency change of $\pm 8\%$, irrespective of the $\Delta M$ value.

\end{itemize}

The limited statistics of the 150 signal samples (1000 events each) and 
the parametrization of the efficiencies with a polynomial are responsible
for 3\% uncertainty. Beam-related backgrounds, 
which affect the determination of $E_{12}$, were not simulated. 
The effect on the 
selection efficiency, determined from events collected at random beam 
crossings, is a relative decrease of 5\%.   

Finally, uncertainties in the simulation of the background were assessed by
comparing the effect of each cut separately on the data and on the simulated 
backgrounds after the preselection. The relative differences, although 
compatible with the uncertainty due to limited statistics, were 
conservatively added in quadrature, for a total possible deviation of 9\%. 
%This possible effect was accounted for by reducing the amount of subtracted 
%background, {\it i.e.}, the four-fermion and the \qqbar\ background, by 9\%.

To derive the final result, all the above uncertainties were taken into
account following the method of Ref.~\cite{cousins}.

\subsubsection{Results and interpretation}

The numbers of candidate events observed in the data between 1997 and 2000 
are displayed in Table~\ref{tab:res}. These numbers are in agreement 
with the numbers of events expected from standard model background sources.
\begin{table}[htbp]
\caption{\footnotesize
The numbers of candidate events observed in the data between 1997 and
2000, and the numbers of events expected from standard model background 
sources, for the small, intermediate and large $\Delta M$ selections.
\label{tab:res}}
\begin{center}
\begin{tabular}{|c|c|c|c|c|c|c|c|}
\hline
\hline
{\hbox{\lower 0.3cm \hbox{Year}}} & 
{\hbox{\lower 0.3cm \hbox{Luminosity [pb$^{-1}$]}}} &
\multicolumn{2}{|c|}{{\hbox{\lower 0.1cm \hbox{Small $\Delta M$}}}} &
\multicolumn{2}{|c|}{{\hbox{\lower 0.1cm \hbox{Int. $\Delta M$}}}} &
\multicolumn{2}{|c|}{{\hbox{\lower 0.1cm \hbox{Large $\Delta M$}}}} \\
\cline{3-8}
\rule{0pt}{14 pt} 
      &       & Obs. & Exp. & Obs. & Exp. & Obs & Exp. \\ \hline\hline
\rule{0pt}{14 pt} 
2000  & 207.3 &  6   &  8.1  & 11   & 14.7 & 17  & 17.0 \\ \hline
\rule{0pt}{14 pt} 
1999  & 236.9 &  9   &  8.7  & 15   & 16.7 & 19  & 19.5 \\ \hline
\rule{0pt}{14 pt} 
1998  & 173.6 &  7   &  6.9  & 11   & 13.4 & 17  & 15.2 \\ \hline
\rule{0pt}{14 pt} 
1997  &  56.8 &  1   &  2.2  &  5   &  4.4 &  4  &  5.0 \\ \hline\hline
\rule{0pt}{14 pt} 
Total & 674.6 & 23   & 25.9  & 42   & 49.2 & 57  & 56.7 \\ \hline\hline
\end{tabular}   
\end{center}
\end{table}

%\begin{table}[htbp]
%\caption{\footnotesize
%The background composition of the small, intermediate and large $\Delta M$ 
%selections. 
%\label{tab:bgcomp}}
%\begin{center}
%\begin{tabular}{|c|c|c|c|c|c|}
%\hline\hline
% Source      & Small $\Delta M$ & Int. $\Delta M$ & Large $\Delta M$ \\
%\hline
%\hline
%\rule{0pt}{14 pt} Two-photon  & 23\%  &  7\%   &   4\%  \\ \hline
%\rule{0pt}{14 pt} \qqbar      &  9\%  &  3\%   &   1\%  \\ \hline
%\rule{0pt}{14 pt} We$\nu$     & 48\%  & 21\%   &  11\%  \\ \hline
%\rule{0pt}{14 pt} WW          & 10\%  & 49\%   &  68\%  \\ \hline
%\rule{0pt}{14 pt} ZZ          & 10\%  & 20\%   &  16\%  \\ \hline
%\hline
%\end{tabular}   
%\end{center}
%\end{table}

\begin{figure}[htbp]
\begin{picture}(160,125)
\put(5,-4){\epsfxsize150mm\epsfbox{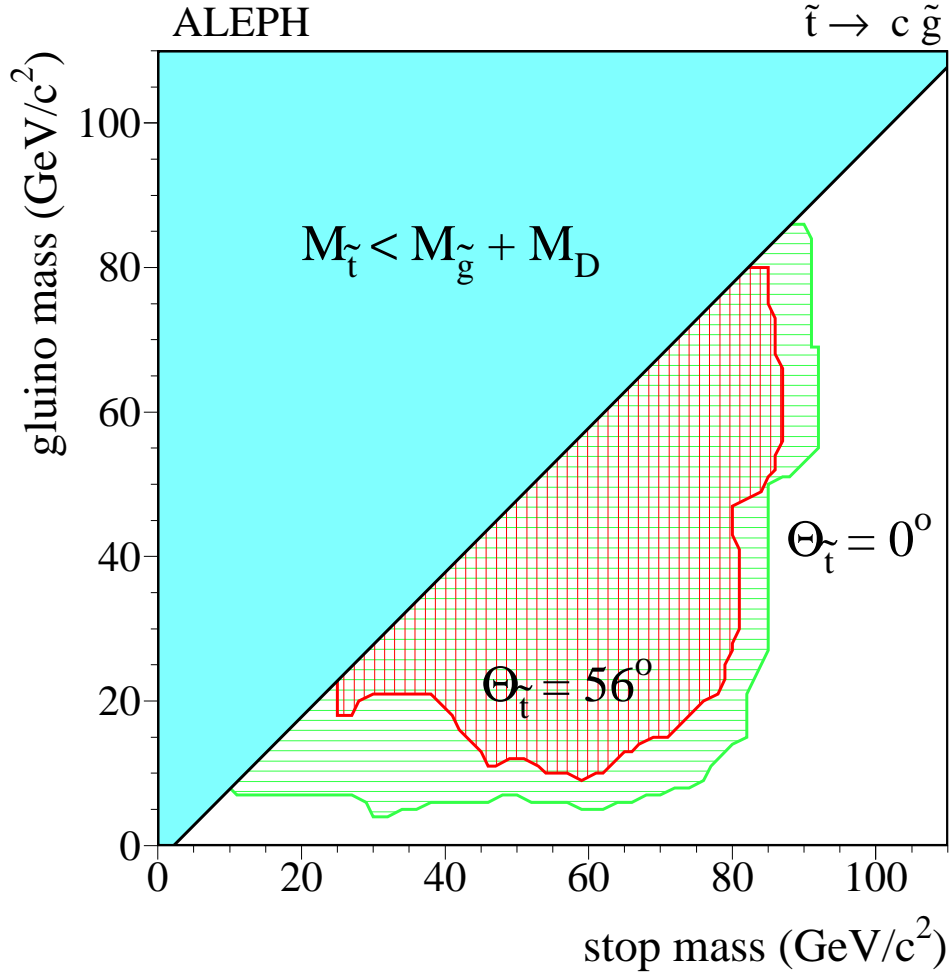}}
\end{picture}
\caption[ ]
{\protect\footnotesize The 95\%\,C.L. excluded regions in the plane 
(\mst, \mglu) from the acoplanar jet search at LEP\,2, for maximal 
(horizontal hatching) and minimal (cross hatching) stop 
coupling to the Z. The shaded area corresponds to a stable stop and 
is not accessible to this search.
\label{fig:excl}}
\end{figure}

In the framework of the MSSM with R-parity conservation,  
the outcome of this search can be translated into
constraints in the (\mst,\mglu) plane when the stop quark 
decays with a 100\% branching fraction into a stable hadronizing 
gluino and a c quark. Regions excluded by this search at 
95\%\,C.L. are shown in Fig.~\ref{fig:excl}, for $\theta_{\rm mix} = 
56^\circ$ and $0^\circ$, corresponding to vanishing and maximal 
stop coupling to the Z, respectively.

\subsection{The {\mbox{\boldmath $\Delta M \simeq m_{\rm D}$}} case} 

The analyses of Sections~6 and~7.1 do not suffice to efficiently cover the 
$\Delta M = m_{\rm D}$ line: along this line, the phase space 
for the $\st \to {\rm c}\glu$ decay is so small that the stop lifetime may 
take any intermediate value, for which none of the previous selections is 
fully effective.

Stop pair production followed by decays into c\glu\ was simulated
for $\Delta M = m_{\rm D}$ as explained in Section~3, with stop proper 
decay lengths varying from 1\,mm to 1\,m. The acoplanar jet search 
of Section~7.1, the search for heavy stable charged particles of Section~6 
and the search for kinks and secondary vertices of Ref.~\cite{Barate:2000qf}
were applied in turn to these simulated samples.
Altogether, they allow all stop masses between 14 and 80\,\gevcc\ 
to be excluded for this very small mass difference, irrespective
of the stop lifetime, for a vanishing stop coupling to the Z.

\section{Combined results}

As can be seen from Fig.~\ref{fig:excl}, no absolute stop mass limit
can be extracted from the results of the acoplanar jet plus missing
energy search alone. An absolute mass limit is obtained by combining
all searches presented in this paper in the following way.
\begin{itemize}

\item Small gluino masses do not give rise to large enough missing energy 
to be addressed by the $\st \to {\rm c}\glu$ search. However, 
the searches for \epemto\ \qqbar\glu\glu\ at LEP\,1 presented in 
Ref.~\cite{janot} and in Section~4 allow all gluino masses below 
26.9\,\gevcc\ to be excluded when the gluino is the LSP ($\mglu < \mst$).

\item Stable stops, {\it i.e.}, with $\Delta M < m_{\rm D}$, are excluded
up to 15.7\,\gevcc\ with the search for \epemto\ \qqbar\sqsqbar\ at LEP\,1 
presented in Section~5, and up to 95\,\gevcc\ by the search for heavy 
stable charged particles presented in Section~6.

\item Promptly decaying heavy stops, {\it i.e.}, with $\Delta M > 
m_{\rm D}$, are excluded up to masses of 85\,\gevcc\ by the acoplanar jet 
search of Section~7.1.

\item For $\Delta M \simeq m_{\rm D}$, masses up to 80\,\gevcc\ are excluded
independently of the stop lifetime, as mentioned in Section~7.2.
\end{itemize}

\begin{figure}[ht]
\begin{picture}(160,145)
\put(0,-4){\epsfxsize170mm\epsfbox{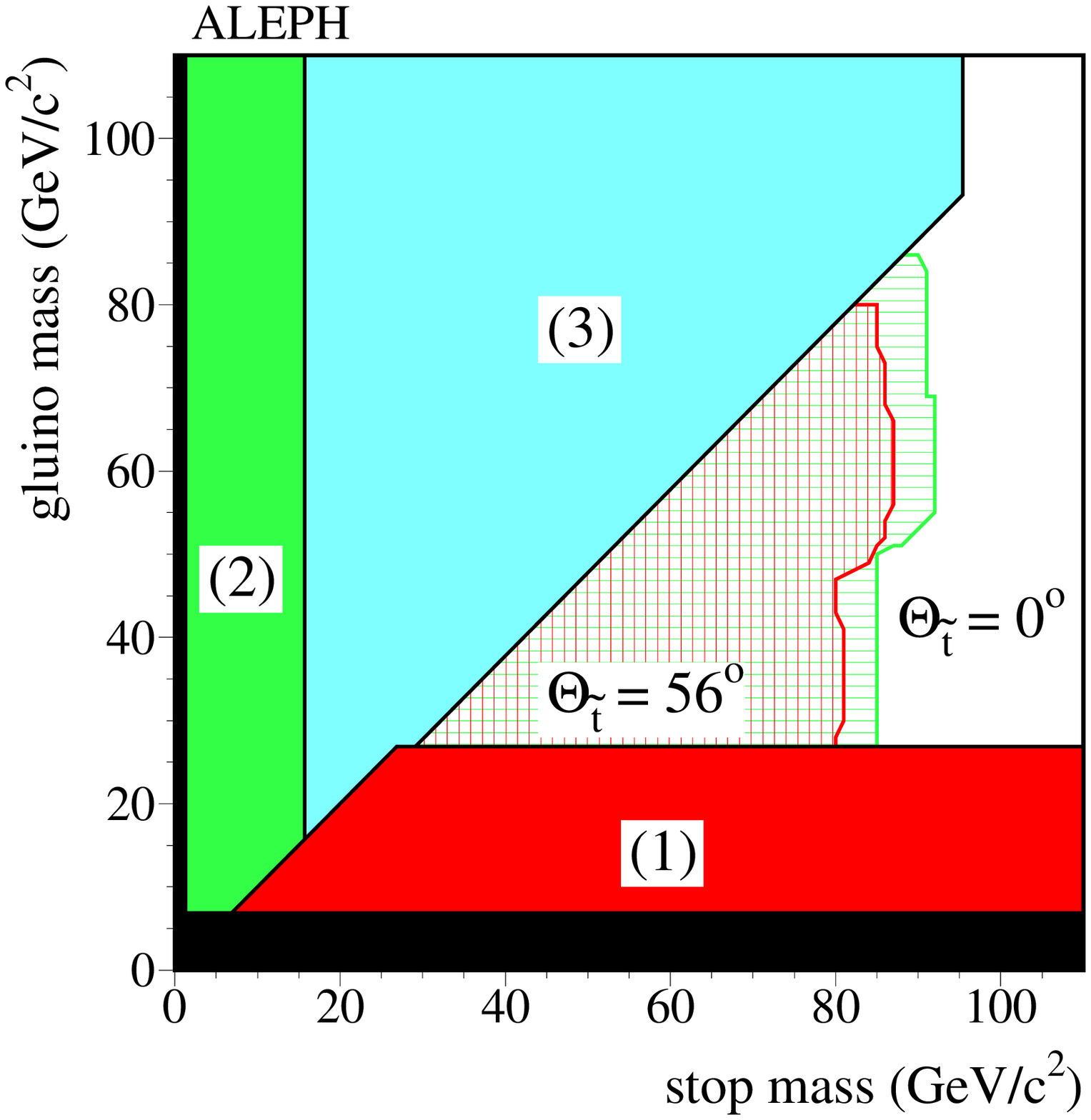}}
\end{picture}
\caption[ ]
{\protect\footnotesize The 95\%\,C.L. excluded regions in the plane 
(\mst, \mglu) from the combination of all searches presented in this
paper. The black area at very small gluino and squark masses is 
excluded by the precise measurement of the Z lineshape; Regions (1), (2) and 
(3) are excluded by the search for \epemto\ \qqbar\glu\glu\ at LEP\,1, for
\epemto\ \qqbar\sqsqbar\ at LEP\,1, and for heavy stable charged particles 
from \sqsqbar\ production at LEP\,2, respectively. The hatched areas are 
excluded by the acoplanar jet with missing energy search at LEP\,2. 
\label{fig:comb}}
\end{figure}

The result of the combination is displayed in Fig.~\ref{fig:comb}.
All stop masses below 80\,\gevcc\ are excluded at 95\%\,C.L. when 
either the stop or the gluino is the lightest supersymmetric 
particle. When combined with the result of Ref.~\cite{squark},
all stop masses below 63\,\gevcc\ are excluded irrespective of the 
nature of the LSP. 

Sbottom quarks were also searched for, but the acoplanar jet search 
proved not to be efficient enough to cope with {\it (i)} the four times 
smaller cross section of $\epemto \sbsbbar\ \to \bbbar\glu\glu$, even 
when b-tagging is applied to reduce the background; and {\it (ii)} the 
very small mass differences (down to $m_{\rm K}$) for which the sbottom may 
decay promptly to s\glu. However, the Z lineshape limits and the excluded
areas (1), (2) and (3) of Fig.~\ref{fig:comb} apply as well for all 
squark species, and in particular for the sbottom.  All sbottom masses below 
27.4\,\gevcc\ are therefore excluded at 95\%\,C.L. when either the 
sbottom or the gluino is the lightest supersymmetric particle.

\eject\null

\section{Conclusions}

Searches for stable, hadronizing squarks and gluinos have been performed 
with the data collected at centre-of-mass energies from 88 to 209\,GeV
by the ALEPH detector at LEP. No evidence for such a signal was observed
in any of the production processes studied: \epemto\ \qqbar\glu\glu, 
\epemto\ \qqbar\sqsqbar\ and \epemto\ \sqsqbar.  The following absolute 
mass limits were obtained at the 95\% confidence level in the framework 
of the MSSM with R-parity conservation:
%\eject
\begin{itemize}
\item a gluino LSP is excluded for $\mglu < 26.9\,\gevcc$;
\item a down-type squark LSP is excluded for $\msq < 92\,\gevcc$;
\item an up-type squark LSP is excluded for $\msq < 95\,\gevcc$;
\item a sbottom quark NSLP is excluded up to masses of 27.4\,\gevcc\ 
if the LSP is a gluino;
\item a stop quark NSLP is excluded up to masses of 80\,\gevcc\ 
if the LSP is a gluino, and up to masses of 63\,\gevcc\ irrespective
of the nature of the LSP.
\end{itemize}

%\eject
The above squark and gluino LSP limits also apply in any supersymmetric
model in which squarks or gluinos are long-lived. In particular, the 
scenario of Ref.~\cite{berger} with a gluino of 12~to~16\,\gevcc\ 
decaying into a b quark and a long-lived sbottom with a mass of 
2 to 5\,\gevcc\ is no longer viable. 

The results presented in this paper improve on related existing 
results~\cite{delphi,delphi2}. In particular, absolute lower limits on 
the masses of stable squarks, stable gluinos, and stop quarks decaying 
into stable gluinos, have been reported for the first time.

\section{Acknowledgements}
We thank our colleagues from the CERN accelerator divisions for the 
successful running of LEP at high energy. We are indebted to the 
engineers and technicians in all our institutions for their contribution 
to the good performance of ALEPH. We would like to thank 
T.\,Sj\"ostrand for his help with the signal simulation. Those 
of us from non-member states thank CERN for its hospitality.

\end{document}